\documentclass[twocolumn]{jpsj3}

\usepackage{txfonts}
\usepackage{graphicx}
\usepackage{dcolumn}
\usepackage{bm}
\usepackage[usenames,dvipsnames]{xcolor}
\usepackage{ulem}
\usepackage{amsmath}
\usepackage{braket}
\usepackage{enumerate}
\usepackage{arydshln}
\usepackage{multirow}

\title{Symmetry Rules on Multipole Interactions under Crystallographic Point Groups and Application to Multiple-$Q$ Multipole States}
\author{Ryota Yambe$^1$, Satoru Hayami$^2$}
\inst{
  $^1$Department of Applied Physics, The University of Tokyo, Tokyo 113-8656, Japan\\
  $^2$Graduate School of Science, Hokkaido University, Sapporo 060-0810, Japan}
\abst{ 
Multipole degrees of freedom describe the mutual interplay among the charge, spin, and orbital degrees of freedom in electrons, which provides a microscopic understanding of unconventional electronic orderings and their associated physical phenomena. 
We here show the symmetry rules on multipole interactions under crystallographic point groups in a systematic manner. 
Depending on the bond symmetries, we show the necessary symmetry conditions of the antisymmetric multipole interactions, which correspond to the extension of the Dzyaloshinskii-Moriya interaction, as well as the symmetric ones, which correspond to the extension of the compasslike interaction. 
Furthermore, we demonstrate that the symmetry-allowed multipole interactions can become a source of exotic multiple-$Q$ multipole orderings. 
As a specific example, we analyze the effective model with the antisymmetric quadrupole interaction on a triangular lattice and show the emergence of the triple-$Q$ quadrupole state. 
Our results indicate that multipole interactions that often arise from the heavy-fermion, frustrated, and nematic systems can potentially induce further unconventional quantum states of matter. 
}

\begin{document}
\maketitle

\section{Introduction}
\label{sec:introduction}

Multipoles are one of the fundamental concepts to describe anisotropic charge and current distributions, which have been long studied in various fields of physics, such as classical electromagnetism~\cite{LandauLifshitz198001, Jackson3rd1999, nanz2016toroidal}, nuclear physics~\cite{BlattWeisskopf201111}, and molecular physics~\cite{stogryn1966molecular}. 
In condensed matter physics, the concept of multipoles has been introduced to describe atomic-scale anisotropic charge and spin distributions arising from the strong spin--orbital entanglement in crystals~\cite{Santini_RevModPhys.81.807, kuramoto2009multipole}. 
Extensive studies have revealed the emergence of higher-rank multipole orderings in a variety of materials, such as rank-2 electric quadrupole orderings in CeB$_6$~\cite{Takigawa_doi:10.1143/JPSJ.52.728,nakao2001antiferro,tanaka2004direct, Portnichenko_PhysRevX.10.021010}, PrPb$_3$~\cite{morin1982magnetic, onimaru2004angle, Onimaru_PhysRevLett.94.197201}, Pr$T_2$$X_{20}$ ($T=$ Ir, Rh, $X=$ Zn; $T=$ V, $X=$ Al)~\cite{Onimaru_PhysRevLett.106.177001, Ishii_doi:10.1143/JPSJ.80.093601, sakai2011kondo, onimaru2016exotic, Ishitobi_PhysRevB.104.L241110}, and Ba$_2$MgReO$_6$
~\cite{Hirai_PhysRevResearch.2.022063, Mansouri_PhysRevMaterials.5.104410, Lovesey_PhysRevB.103.235160}, rank-3 magnetic octupole orderings in Ce$_{1-x}$La$_{x}$B$_6$~\cite{Mannix_PhysRevLett.95.117206}, and rank-4 electric hexadecapole orderings in PrRu$_4$P$_{12}$~\cite{lee2001structural,takimoto2006antiferro}, which exhibit qualitatively different properties from conventional rank-1 magnetic and electric dipole orderings. 
Furthermore, toroidal-type multipole orderings have also been found or proposed in both magnetic and nonmagnetic materials, such as rank-0 magnetic toroidal monopole ordering in Co$_2$SiO$_4$~\cite{Hayami_PhysRevB.108.L140409, hayashida2024electric}, rank-0 electric toroidal monopole ordering in URu$_2$Si$_2$~\cite{Kambe_PhysRevB.97.235142, kambe2020symmetry, hayami2023chiral}, rank-1 magnetic toroidal dipole ordering in Cr$_2$O$_3$~\cite{FolenPhysRevLett.6.607, popov1999magnetic, krotov2001magnetoelectric}, rank-1 electric toroidal dipole ordering in RbFe(MoO$_4$)$_2$~\cite{Hlinka_PhysRevLett.116.177602, jin2020observation, Hayashida_PhysRevMaterials.5.124409, Hayami_doi:10.7566/JPSJ.91.113702, cheong2021permutable}, and rank-2 electric toroidal quadrupole ordering in Cd$_2$Re$_2$O$_7$~\cite{yamaura2002low, hiroi2018pyrochlore, Matteo_PhysRevB.96.115156, Hayami_PhysRevLett.122.147602}. 

In the early stage of the multipole investigations, simple alignments of such higher-rank multipoles have been mainly investigated, as found in CeAg~\cite{ray1976occurrence, morin1988quadrupolar}, TmZn~\cite{Morin_PhysRevB.14.2972, luthi1979quadrupolar}, and K$_2$CuF$_4$~\cite{kubo1976systematic}. 
Meanwhile, recent studies show more exotic multipole orderings, such as multiple-$Q$ quadrupole or dipole-quadrupole orderings~\cite{tsunetsugu2021quadrupole, Ishitobi_PhysRevB.104.L241110, Yanagisawa_PhysRevLett.126.157201, Ishitobi_PhysRevB.107.104413, Remund_PhysRevResearch.4.033106, Seifert_PhysRevB.106.195147, Hattori_PhysRevB.107.205126, hayami2023multipleJPSJ, Hattori_PhysRevResearch.6.L042068}. 
In addition, topologically nontrivial multiple-$Q$ multipole orderings, such as the CP(2) skyrmion consisting of multiple-$Q$ dipole and quadrupole density waves, have been revealed in the context of nontrivial low-energy excitations~\cite{mikushina2002dipole, Garaud_PhysRevB.87.014507, Akagi_PhysRevD.103.065008, Amari_PhysRevB.106.L100406, zhang2023cp2, Benfenati_PhysRevB.107.094503, Zhang_PhysRevLett.133.196702}. 
Meanwhile, stabilization mechanisms of these multiple-$Q$ multipole orderings have not been fully elucidated. 
This is because the emergence of multiple-$Q$ multipole orderings depends on microscopic interactions between higher-rank multipoles, and the types of such interactions are large compared to the types of interactions between dipoles.
Therefore, it is highly desired to systematically obtain their symmetry conditions under crystallographic point groups. 
Such a systematic derivation would be helpful to search for further unknown multiple-$Q$ multipole orderings in future studies.

In the present study, we investigate the symmetry rules on multipole interactions under crystallographic point groups, which have been often studied for specific situations~\cite{sakai2003invariant, Kim_PhysRevB.87.205119, hattori2014antiferro}. 
By using group theory, we classify all the multipole interactions, which include both symmetric and antisymmetric components, according to the bond symmetry. 
The results provide several tables for different ranks of multipoles and different crystallographic point group symmetries, which complete all the possibilities of multipole interactions up to rank 4 in crystals. 
Based on the symmetry classification, we construct an effective model with an antisymmetric electric quadrupole interaction on a two-dimensional triangular lattice under the $C_{\rm 6v}$ symmetry and calculate the ground-state quadrupole configuration. 
Consequently, we show that a triple-$Q$ quadrupole state becomes the ground state by the competition between the isotropic and antisymmetric quadrupole interactions. 
The present scheme enables us to construct effective multipole models with various types of interactions in various lattice structures, which will be a reference to explore stabilization mechanisms of multiple-$Q$ multipole orderings. 

The rest of this paper is organized as follows: 
In Sec.~\ref{sec:int}, we show the symmetry rules on multipole interactions and classify them under crystallographic point groups. 
Then, we apply the classification results to the electric quadrupole system under noncentrosymmetric point group $C_{\rm 6v}$ in Sec.~\ref{sec:Class_2}. 
We demonstrate that the antisymmetric quadrupole interaction gives rise to the triple-$Q$ quadrupole state. 
Section~\ref{sec:summary} is devoted to the summary of the present paper. 
In Appendix~\ref{sec: app1}, we show the classification of multipole interactions under the cubic bases for the [100] bond. 
In Appendices~\ref{sec: app2} and \ref{sec: app3}, we classify the multipole interactions under the hexagonal and cubic bases for the [001] bond, respectively.

\section{Classification of Multipole Interactions}
\label{sec:int}

We present the symmetry rules on multipole interactions under crystallographic point groups. 
In Sec.~\ref{sec:int_def}, we introduce the multipole moments and their interactions. 
Then, we show the representation of the multipole interaction based on the group theory in Sec.~\ref{sec:int_method}. 
We classify the interactions for the dipole, quadupole, octupole, and hexadecapole components in Secs.~\ref{sec:int_dipole}, \ref{sec:int_quadrupole}, \ref{sec:int_octupole}, and \ref{sec:int_hexadecapole}, respectively. 
In these sections, we focus on the classification of the multipole interaction with the hexagonal bases for the [100] bond. 
We present the result with the cubic bases for the [100] bond in Appendix~\ref{sec: app1}. 
We also present the result with hexagonal and cubic bases for the [001] bond in Appendices~\ref{sec: app2} and \ref{sec: app3}, respectively.

\subsection{Definition of Multipole Interactions}
\label{sec:int_def}

We introduce the multipoles in atomic scales~\cite{Kusunose_JPSJ.77.064710, hayami2018microscopic}. 
The multipoles at site $i$ are given by
\begin{align}
\label{eq:multipole_def}
\mathcal{O}^{lm}_i=\sqrt{\frac{4\pi}{2l+1}}|\bm{r}_i|^lY^*_{lm}(\hat{\bm{r}}_i),
\end{align}
where $\bm{r}_i=\bm{r}-\bm{R}_i$ is the position of the electron from the atomic site $\bm{R}_i$, and $Y_{lm}(\hat{\bm{r}}_i)=(-1)^mY^*_{lm}(\hat{\bm{r}}_i)$ with $\hat{\bm{r}}_i=\bm{r}_i/|\bm{r}_i|$ is the spherical harmonics with the azimuthal quantum number of $l$ and the magnetic quantum number of $m$ $(-l\le m \le l)$.
Equation~(\ref{eq:multipole_def}) represents multipoles with rank $l$: monopole ($l=0$), dipole ($l=1$), quadrupole ($l=2$), octupole ($l=3$), hexadecapole ($l=4$), and so on.
The rank-$l$ multipoles are the basis of the $(2l+1)$-dimensional irreducible representations for a full rotational group.

In crystals with discrete rotational symmetries, the $(2l+1)$-fold degeneracy of the rank-$l$ multipoles is lifted by a reduction from the full rotational group to the crystallographic point groups. 
The basis of each irreducible representation can be given by multipoles within real expressions, which depend on the crystallographic point group. 
For instance, we give such real bases of the multipoles up to $l=4$ for the hexagonal group $6/mmm$ with the irreducible representation and its subgroups in Table~\ref{tab:hexagonal_basis}: monopole  $M^\alpha$ with $\alpha=1$, dipole  $D^\alpha$ with $\alpha=1$-3, quadrupole  $Q^\alpha$ with $\alpha=1$-5, octupole  $O^\alpha$ with $\alpha=1$-7, and hexadecapole  $H^\alpha$ with $\alpha=1$-9.
The real bases for the cubic point group $m\bar{3}m$ and its subgroups are also shown in Table~\ref{tab:cubic_basis} in Appendix~\ref{sec: app1}.
In the following, $\mathcal{O}^{\alpha}_i$ with $\alpha=1$-$(2l+1)$ represent the rank-$l$ multipoles with the real expressions.

The rank-$l$ multipoles are further classified into four types according to the spatial inversion ($\mathcal{P}$) and time-reversal ($\mathcal{T}$) parities~\cite{dubovik1975multipole, dubovik1986axial, dubovik1990toroid, kopaev2009toroidal, Spaldin_0953-8984-20-43-434203, hayami2018microscopic, Hayami_PhysRevB.98.165110, talebi2018theory, Yatsushiro_PhysRevB.105.155157, kusunose2022generalization, hayami2024unified}: electric multipoles with $(\mathcal{P},\mathcal{T})=[(-1)^l,+1]$, magnetic multipoles with $(\mathcal{P},\mathcal{T})=[(-1)^{l+1},-1]$, magnetic toroidal multipoles with $(\mathcal{P},\mathcal{T})=[(-1)^{l},-1]$, and electric toroidal multipoles with $(\mathcal{P},\mathcal{T})=[(-1)^{l+1},+1]$, where $+1$ ($-1$) denotes the even (odd) parity. 
These multipoles constitute a complete basis set in physical space, irrespective of atomic scale~\cite{hayami2018microscopic, kusunose2020complete} and cluster scale~\cite{Kusunose_PhysRevB.107.195118}.

\begin{table*}
\centering
\caption{\label{tab:hexagonal_basis}
Hexagonal bases of the monopole $M^\alpha$, dipoles $D^\alpha$, quadrupoles  $Q^\alpha$, octupoles  $O^\alpha$, and  hexadecapoles  $H^\alpha$.
$\hat{\bm{r}}_i=(x_i,y_i,z_i)$ is the unit vector. 
The irreducible representations (Irrep.) of electric and magnetic toroidal multipoles under $D_{\rm 6h}$ point group, except for the time-reversal property, are also shown; in the case of electric toroidal and magnetic multipoles, the subscript $g/u$ is replaced with $u/g$; see Ref.~\citen{Yatsushiro_PhysRevB.104.054412} for the specific expressions of other multipoles.
}
\begin{tabular}{ccc}
\hline\hline
rank $l$ & Hexagonal bases & Irrep. \\ \hline
$0$ & $M_i^1=1$ & $A_{1g}$\\
$1$ & $D_i^1=x_i$ $D_i^2=y_i$ & $E_{1u}$ \\
&  $D_i^3=z_i$ & $A_{2u}$\\
$2$ & $Q_i^1=\dfrac{1}{2}(3z_i^2-r_i^2)$ & $A_{1g}$ \\
&           $Q_i^2=\sqrt{3}z_ix_i$, $Q_i^3=\sqrt{3}y_iz_i$ & $E_{1g}$ \\
&  $Q_i^4=\sqrt{3}x_iy_i$, $Q_i^5=\dfrac{\sqrt{3}}{2}(x_i^2-y_i^2)$ & $E_{2g}$  \\
$3$ & $O_i^1=\dfrac{1}{2}z_i(5z_i^2-3r_i^2)$ & $A_{2u}$ \\
&  $O_i^2=\dfrac{\sqrt{10}}{4}x_i(x_i^2-3y_i^2)$ & $B_{1u}$ \\
&  $O_i^3=\dfrac{\sqrt{10}}{4}y_i(3x_i^2-y_i^2)$ & $B_{2u}$ \\ 
&$O_i^4=\dfrac{\sqrt{6}}{4}x_i(5z_i^2-r_i^2)$,  $O_i^5=\dfrac{\sqrt{6}}{4}y_i(5z_i^2-r_i^2)$ & $E_{1u}$ \\
& $O_i^6=\sqrt{15}x_iy_iz_i$, $O_i^7=\dfrac{\sqrt{15}}{2}z_i(x_i^2-y_i^2)$ & $E_{2u}$ \\
$4$ & $H_i^1=\dfrac{1}{8}\left(35z_i^4-30z_i^2r_i^2+3r_i^4\right)$ & $A_{1g}$ \\
 & $H_i^2=\dfrac{\sqrt{70}}{4}y_iz_i(3x_i^2-y_i^2) $ & $B_{1g}$ \\
& $H_i^3=\dfrac{\sqrt{70}}{4}z_ix_i(x_i^2-3y_i^2) $ & $B_{2g}$\\ 
&$H_i^4=\dfrac{\sqrt{10}}{4}z_ix_i(7z_i^2-3r_i^2)$, $H_i^5=\dfrac{\sqrt{10}}{4}y_iz_i(7z_i^2-3r_i^2)$ & $E_{1g}$ \\
& $H_i^6=\dfrac{\sqrt{35}}{8}(x_i^4-6x_i^2y_i^2+y_i^4)$, $H_i^7=\dfrac{\sqrt{35}}{2}x_iy_i(x_i^2-y_i^2)$  & $E_{2g}$ \\ 
& $H_i^8=\dfrac{\sqrt{5}}{2}x_iy_i(7z_i^2-r_i^2)$, $H_i^9=\dfrac{\sqrt{5}}{4}(x_i^2-y_i^2)(7z_i^2-r_i^2)$ & $E_{2g}$
\\ \hline\hline
\end{tabular}
\end{table*}

Let us consider a bilinear interaction between the rank-$l$ multipoles at different sites $i$ and $j$, which are written in the form of
\begin{align}
\label{eq:multiple_int}
\sum_{\alpha\beta} J^{\alpha\beta}_{ij}\mathcal{O}^{\alpha}_i\mathcal{O}^{\beta}_j,
\end{align}
where $\mathcal{O}^{\alpha}_i$ is the $\alpha=1,\cdots,2l+1$ component of the rank-$l$ multipole at site $i$ and $J^{\alpha\beta}_{ij}$ is the real coupling constant.
The interaction matrix $J_{ij}$ corresponds to the $(2l+1)\times (2l+1)$ matrix, which has $(2l+1)^2$ real components. 
According to the parity with respect to the interchange of the multipole components, each component is classified into two types according to symmetric matrix components $J^{\mathrm{(S)}\alpha\beta}_{ij}=(J^{\alpha\beta}_{ij}+J^{\beta\alpha}_{ij})/2$ and antisymmetric matrix components $J^{\mathrm{(AS)}\alpha\beta}_{ij}=(J^{\alpha\beta}_{ij}-J^{\beta\alpha}_{ij})/2$.
 
Then, the bilinear multipole interactions in Eq.~(\ref{eq:multiple_int}) are divided into three types of interactions:
\begin{align}
&J^{\rm iso}_{ij} \sum_{\alpha=1}^{2l+1}\mathcal{O}_i^\alpha\mathcal{O}_j^\alpha, \\
&(J^{\mathrm{(S)}\alpha\beta}_{ij} -\delta_{\alpha\beta}J^{\rm iso}_{ij}) (\mathcal{O}^{\alpha}_i\mathcal{O}^{\beta}_j+\mathcal{O}^{\beta}_i\mathcal{O}^{\alpha}_j) ,\\
&J^{\mathrm{(A S)}\alpha\beta}_{ij} (\mathcal{O}^{\alpha}_i\mathcal{O}^{\beta}_j-\mathcal{O}^{\beta}_i\mathcal{O}^{\alpha}_j). 
\end{align}
The first is an isotropic interaction with $J^{\rm iso}_{ij} = \mathrm{Tr}[J_{ij}]/(2l+1)$, the second is a symmetric anisotropic interaction, and the third is an antisymmetric anisotropic interaction. 
The symmetric and antisymmetric anisotropic interactions are also characterized by even and odd parities with respect to the interchange of sites, respectively. 
It is noted that the magnitude of the higher-rank multipole interaction, such as the electric quadrupole interaction, can be comparable to that of the dipole interaction depending on electronic structures~\cite{ohkawa1983ordered, shiina1997magnetic, yamada2019derivation, Otsuki_PhysRevB.110.035104}. 

\subsection{Method}
\label{sec:int_method}

The form of the multipole interactions in Eq.~(\ref{eq:multiple_int}) is determined to satisfy the symmetry of the $\langle i,j \rangle$ bond.
In the case of the monopole $(l=0)$, the interaction matrix has a single component $J^{\mathrm{iso}}_{ij}$, which is allowed independent of the point group.
In contrast, nonzero components in $J_{ij}$ for $l\ge1$ depend on the point group symmetry.
Such nonzero components of the multipole interactions can be obtained by using the representation theory for the multipoles at two sites, as follows. 
For that purpose, we rewrite the interactions in Eq.~(\ref{eq:multiple_int}) as~\cite{Yambe_PhysRevB.108.064420}  
\begin{align}
\sum_{\alpha,\beta}\left( J^{\alpha\beta}_{ij}\mathcal{O}^{\alpha}_i\mathcal{O}^{\beta}_j +J^{\alpha\beta}_{ji}\mathcal{O}^{\alpha}_j\mathcal{O}^{\beta}_i \right)=
\tilde{\mathcal{O}}^{\top}
\begin{pmatrix}
0 & J_{ij} \\
 J_{ij}^{\top} & 0
\end{pmatrix}
\tilde{\mathcal{O}},
\end{align}
where $\tilde{\mathcal{O}}^{\top}=(\mathcal{O}^1_{i},\cdots,\mathcal{O}^{2l+1}_{i},\mathcal{O}^1_{j},\cdots,\mathcal{O}^{2l+1}_{j})$ and we use a relation of $J_{ji} = J_{ij}^{\top}$; $\top$ represents the transpose.
When the bond symmetry is preserved by a point group operation $R$, the interaction matrix satisfies the symmetry constraint given by
\begin{align}
\label{eq:constraint}
\begin{pmatrix}
0 & J_{ij} \\
 J_{ij}^{\top} & 0
\end{pmatrix} =  \Gamma^{\top}(R)
\begin{pmatrix}
0 & J_{ij} \\
 J_{ij}^{\top} & 0
\end{pmatrix} \Gamma(R), 
\end{align}
where $\Gamma(R)=\Gamma_\mathrm{perm}(R)\otimes\Gamma_l (R)$ is the representation with respect to the point group operation $R$; the $2\times 2$ matrix $\Gamma_\mathrm{perm}(R)$ is a permutation representation for the sites $i$ and $j$, and the $(2l+1)\times (2l+1)$ matrix $\Gamma_l(R)$ represents the transformation of the rank-$l$ multipoles.
$\Gamma_\mathrm{perm}(R)$ and $\Gamma_l(R)$ are obtained from
 \begin{align}
 \label{eq:gamma_perm}
R(i,j)^{\top}&=\Gamma_\mathrm{perm}(R)(i,j)^{\top}, \\
\label{eq:gamma_l}
R(\mathcal{O}^{1},\cdots,\mathcal{O}^{2l+1})^{\top}&=\Gamma_{l}(R)(\mathcal{O}^{1},\cdots,\mathcal{O}^{2l+1})^{\top},
 \end{align}
respectively.    
The form of $\Gamma_{\rm perm}(R)$ is independent of the type of multipoles, while the form of $\Gamma_{l}(R)$ depends on the type of multipoles.

The point group operation preserving the bond symmetry is classified into two types based on the transformation of the bond: type-I operation $R^\mathrm{I}$ satisfying $R^\mathrm{I}(i, j)^{\top} = (i, j)^{\top}$ and type-II operation $R^\mathrm{II}$ satisfying $R^\mathrm{II}(i, j)^{\top} = (j, i)^{\top}$.
Thus, the permutation representations for $R^\mathrm{I}$ and $R^\mathrm{II}$ are explicitly given by 
\begin{align}
\Gamma_\mathrm{perm}^\mathrm{I}&=
\begin{pmatrix}
1 & 0\\
0 & 1
\end{pmatrix},\\
\Gamma_\mathrm{perm}^\mathrm{II}&=
\begin{pmatrix}
0 & 1\\
1 & 0
\end{pmatrix},
\end{align}
respectively.
Although we only consider the case of the real-space interaction, the above argument is straightforwardly applied to the momentum-resolved multipole interaction as studied for the spin model~\cite{Yambe_PhysRevB.106.174437, Yambe_PhysRevB.107.174408}.

\subsection{Dipole Interactions}
\label{sec:int_dipole}

\begin{table*}
\centering
\caption{
\label{tab:dipole_hexa_100}
Classification of the dipole interaction with the hexagonal bases for the $[100]$ bond.
The components with $\checkmark$ (0) are symmetry allowed (not allowed).
The diagonal components $J^{\mathrm{(S)}\alpha\alpha}_{ij}$ ($\alpha=1,2,3$), which are omitted in the table, are symmetry allowed irrespective of the point groups.  
The 1st, 2nd, and 3rd axes in the point group are set to $[001]$, $[100]$, and $[010]$, respectively. 
The row ``\#" represents the number of independent interaction parameters except for the diagonal components.
}
\scalebox{1.0}{
\begin{tabular}{lccccccccccccccc}
\hline \hline 
& $mmm$ & $2mm$ & $m2m$ & $mm2$ & $222$ & $2/m..$ & $.2/m.$ & $..2/m$ & $2..$ & $.2.$ & $..2$ & $m..$ & $.m.$ & $..m$ & $\bar{1}$   \\ \hline
$J^{\mathrm{(AS)}12}_{ij}$&0 &0 &0	 &$\checkmark$	&0 &0 &0 &0	 &0 &0 &$\checkmark$ &$\checkmark$ &$\checkmark$ &0 &0\\
$J^{\mathrm{(AS)}13}_{ij}$&0 &$\checkmark$ &0 &0	&0 &0 &0 &0	 &$\checkmark$ &0 &0 &0 &$\checkmark$ &$\checkmark$ &0\\
$J^{\mathrm{(AS)}23}_{ij}$&0 &0 &0 &0	&$\checkmark$ &0 &0 &0 &$\checkmark$ &$\checkmark$ &$\checkmark$ &0 &0 &0 &0\\
$J^{\mathrm{(S)}12}_{ij}$  &0 &0 &0	 &0	&0 &$\checkmark$ &0 &0	 &$\checkmark$ &0 &0 &$\checkmark$ &0 &0 &$\checkmark$\\
$J^{\mathrm{(S)}13}_{ij}$  &0 &0 &0	 &0	&0 &0 &0 &$\checkmark$	 &0 &0 &$\checkmark$ &0 &0 &$\checkmark$ &$\checkmark$\\
$J^{\mathrm{(S)}23}_{ij}$  &0 &0 &0	 &0	&0 &0 &$\checkmark$ &0	 &0 &$\checkmark$ &0 &0 &$\checkmark$ &0 &$\checkmark$
\\ \hline
\# & 0 &  1 & 0 & 1 & 1 & 1 & 1 & 1 & 3 & 2 & 3 & 2 & 3 & 2 & 3
\\ \hline \hline 
\end{tabular}
}
\end{table*}

We classify the dipole interactions for the bond along the $[100]$ direction, where we use the hexagonal bases.
In the hexagonal systems, the symmetry of the $[100]$ bond is classified into 16 point groups: $mmm$, $2mm$, $m2m$, $mm2$, $222$, $2/m..$, $.2/m.$, $..2/m$, $2..$, $.2.$, $..2$, $m..$, $.m.$, $..m$, $\bar{1}$, and 1, where the 1st, 2nd, and 3rd axes are $[001]$, $[100]$, and $[010]$, respectively. 
These point groups consist of the identity, the spatial inversion ($I$), mirror perpendicular to the $\xi=x,y,z$ axis ($m_{\xi}$), and twofold rotation around the $\xi$ axis ($C_{2\xi}$). 
Following Eqs.~(\ref{eq:gamma_perm}) and (\ref{eq:gamma_l}), we obtain their representations as  
\begin{align}
\label{eq:dipole_I}
\Gamma(I)&=\Gamma_\mathrm{perm}^\mathrm{II}\otimes
\mathcal{P}\begin{pmatrix}
1 & 0 & 0\\
0 & 1 & 0\\
0 & 0 & 1
\end{pmatrix},\\
\label{eq:dipole_mx}
\Gamma(m_x)&=\Gamma_\mathrm{perm}^\mathrm{II}\otimes
\mathcal{P}\begin{pmatrix}
1 & 0 & 0\\
0 & -1 & 0\\
0 & 0 & -1
\end{pmatrix},\\
\Gamma(m_y)&=\Gamma_\mathrm{perm}^\mathrm{I}\otimes
\mathcal{P}\begin{pmatrix}
-1 & 0 & 0\\
0 & 1 & 0\\
0 & 0 & -1
\end{pmatrix},\\
\label{eq:dipole_mz}
\Gamma(m_z)&=\Gamma_\mathrm{perm}^\mathrm{I}\otimes
\mathcal{P}\begin{pmatrix}
-1 & 0 & 0\\
0 & -1 & 0\\
0 & 0 & 1
\end{pmatrix},\\
\Gamma(C_x)&=\Gamma_\mathrm{perm}^\mathrm{I}\otimes
\begin{pmatrix}
1 & 0 & 0\\
0 & -1 & 0\\
0 & 0 & -1
\end{pmatrix},\\
\Gamma(C_y)&=\Gamma_\mathrm{perm}^\mathrm{II}\otimes
\begin{pmatrix}
-1 & 0 & 0\\
0 & 1 & 0\\
0 & 0 & -1
\end{pmatrix},\\
\label{eq:dipole_Cz}
\Gamma(C_z)&=\Gamma_\mathrm{perm}^\mathrm{II}\otimes
\begin{pmatrix}
-1 & 0 & 0\\
0 & -1 & 0\\
0 & 0 & 1
\end{pmatrix},
\end{align}
where $\mathcal{P}=\pm1$ represents the spatial inversion parity of the dipoles; we set $\mathcal{P}=-1$ ($+1$) for the electric dipoles and magnetic toroidal dipoles (magnetic dipoles and electric toroidal dipoles).
The symmetry constraints are obtained by substituting Eqs.~(\ref{eq:dipole_I})-(\ref{eq:dipole_Cz}) into Eq.~(\ref{eq:constraint}).
It is noted that the same symmetry constraints are obtained for electric, magnetic, magnetic toroidal, and electric toroidal dipoles independently of their types, since the spatial inversion parity $\mathcal{P}$ in Eqs.~(\ref{eq:dipole_I})-(\ref{eq:dipole_Cz}) are canceled out in Eq.~(\ref{eq:constraint}).

We show the classification of the dipole interactions with the hexagonal bases shown in Table \ref{tab:hexagonal_basis} for the [100] bond in Table~\ref{tab:dipole_hexa_100}. 
The results show that the diagonal components of the symmetric anisotropic interactions are allowed irrespective of the symmetry of the bond, while the nonzero off-diagonal components of the symmetric and antisymmetric anisotropic interactions depend on the symmetry of the bonds. 
The inversion symmetry forbids (allows) the off-diagonal antisymmetric (symmetric) anisotropic interactions, as shown in the result for $\bar{1}$.
The mirror symmetries impose the symmetry constraints depending on the mirror planes, as shown in the results for $m..$, $.m.$, and $..m$.
By comparing the symmetry constraints on the off-diagonal antisymmetric and symmetric anisotropic interactions,  
 $m_z$ and $m_y$ allow the same components, while the $m_x$ allows the different components; $m_z$ ($m_y$) allows $J^{\mathrm{(AS)}12}_{ij}$ and $J^{\mathrm{(S)}12}_{ij}$ ($J^{\mathrm{(AS)}13}_{ij}$ and $J^{\mathrm{(S)}13}_{ij}$), while $m_x$ allows $J^{\mathrm{(AS)}12}_{ij}$, $J^{\mathrm{(AS)}13}_{ij}$ and $J^{\mathrm{(S)}23}_{ij}$.
Such a difference between the symmetry constraints by ($m_z,m_y$) and $m_x$ appears because $m_z$ and $m_y$ are the type-I operations leaving the sites invariant, while $m_x$ is the type-II operation interchanging the sites.
The twofold rotational symmetries impose the symmetry constraints depending on the rotational axes, as shown in the results for $2..$, $.2.$, and $..2$.  
The type-I operation $C_x$ allows the same components of the off-diagonal antisymmetric and symmetric anisotropic interactions, while the type-II operations $C_z$ and $C_y$ allow the different components.
Results for other point groups are obtained by combining the results for $\bar{1}$, $m_x$, $m_y$, $m_z$, $C_x$, $C_y$, and $C_z$.

In the case of the magnetic dipoles (spins), the classification in Table~\ref{tab:dipole_hexa_100} corresponds to the classification of bilinear magnetic (spin) interactions: isotropic exchange interaction, symmetric anisotropic exchange interactions~\cite{kaplan1983single, jackeli2009mott}, and antisymmetric anisotropic exchange interactions called the Dzyaloshinskii-Moriya (DM) interactions~\cite{dzyaloshinsky1958thermodynamic,moriya1960anisotropic}.
These bilinear magnetic interactions become a source of the noncoplanar spin textures, such as the skyrmion, hedgehog, and vortex crystals~\cite{rossler2006spontaneous, Yi_PhysRevB.80.054416, Mochizuki_PhysRevLett.108.017601, amoroso2020spontaneous, Hayami_PhysRevB.103.054422, yambe2021skyrmion, Wang_PhysRevB.103.104408, Kato_PhysRevB.104.224405, amoroso2021tuning, hayami2024stabilization}.
The results in Table~\ref{tab:dipole_hexa_100} also show that the symmetry allows not only the bilinear interactions of magnetic dipoles but also the bilinear interactions of electric dipoles, magnetic toroidal dipoles, and electric toroidal dipoles.
This would indicate the possibility of noncoplanar textures of electric dipoles, magnetic toroidal dipoles, and electric toroidal dipoles by the bilinear interactions, such as the electric (polar) skyrmions~\cite{Hong_PhysRevB.81.172101, gregg2012exotic, Thorner_PhysRevB.89.220103, nahas2015discovery, das2019observation, pereira2019theoretical, McCarter_PhysRevLett.129.247601}.

\subsection{Quadrupole Interactions}
\label{sec:int_quadrupole}

\begin{table*}
\centering
\caption{
Classification of the antisymmetric quadrupole interaction with the hexagonal bases for the $[100]$ bond.
The row ``\#" represents the number of the independent interaction parameters.
\label{tab:DM_Q}
}
\scalebox{1.0}{
\begin{tabular}{cccccccccccccccc}
\hline \hline 
& $mmm$ & $2mm$ & $m2m$ & $mm2$ & $222$ & $2/m..$ & $.2/m.$ & $..2/m$ & $2..$ & $.2.$ & $..2$ & $m..$ & $.m.$ & $..m$ & $\bar{1}$   \\ \hline
$J^{\mathrm{(AS)}12}_{ij}$&0	&$\checkmark$&0	&0	&0	&0	&0	&0	&$\checkmark$&0	&0	&0	&$\checkmark$&$\checkmark$&0\\
$J^{\mathrm{(AS)}13}_{ij}$&0	&0	&0	&0	&$\checkmark$&0	&0	&0	&$\checkmark$&$\checkmark$&$\checkmark$&0	&0	&0	&0\\
$J^{\mathrm{(AS)}14}_{ij}$&0	&0	&0	&$\checkmark$&0	&0	&0	&0	&0	&0	&$\checkmark$&$\checkmark$&$\checkmark$&0	&0\\
$J^{\mathrm{(AS)}15}_{ij}$&0	&0	&$\checkmark$&0	&0	&0	&0	&0	&0	&$\checkmark$&0	&$\checkmark$&0	&$\checkmark$&0\\
$J^{\mathrm{(AS)}23}_{ij}$&0	&0	&0	&$\checkmark$&0	&0	&0	&0	&0	&0	&$\checkmark$&$\checkmark$&$\checkmark$&0	&0\\
$J^{\mathrm{(AS)}24}_{ij}$&0	&0	&0	&0	&$\checkmark$&0	&0	&0	&$\checkmark$&$\checkmark$&$\checkmark$&0	&0	&0	&0\\
$J^{\mathrm{(AS)}25}_{ij}$&0	&$\checkmark$&0	&0	&0	&0	&0	&0	&$\checkmark$&0	&0	&0	&$\checkmark$&$\checkmark$&0\\
$J^{\mathrm{(AS)}34}_{ij}$&0	&$\checkmark$&0	&0	&0	&0	&0	&0	&$\checkmark$&0	&0	&0	&$\checkmark$&$\checkmark$&0\\
$J^{\mathrm{(AS)}35}_{ij}$&0	&0	&0	&0	&$\checkmark$&0	&0	&0	&$\checkmark$&$\checkmark$&$\checkmark$&0	&0	&0	&0\\
$J^{\mathrm{(AS)}45}_{ij}$&0	&0	&0	&$\checkmark$&0	&0	&0	&0	&0	&0	&$\checkmark$&$\checkmark$&$\checkmark$&0	&0 
\\ \hline
\# & 0 &  3 & 1 & 3 & 3 & 0 & 0 & 0 & 6 & 4 & 6 & 4 & 6 & 4 & 0
\\ \hline \hline 
\end{tabular}
}
\end{table*}

\begin{table*}
\centering
\caption{
Classification of the symmetric quadrupole interaction with the hexagonal bases for the $[100]$ bond.
The row ``\#" represents the number of the independent interaction parameters except for the diagonal components.
\label{tab: Sym_Q}
}
\scalebox{1.0}{
\begin{tabular}{cccccccccccccccc}
\hline \hline 
& $mmm$ & $2mm$ & $m2m$ & $mm2$ & $222$ & $2/m..$ & $.2/m.$ & $..2/m$ & $2..$ & $.2.$ & $..2$ & $m..$ & $.m.$ & $..m$ & $\bar{1}$   \\ \hline
$J^{\mathrm{(S)}12}_{ij}$&0	&0	&0	&0	&0	&0	&0	&$\checkmark$&0	&0	&$\checkmark$&0	&0	&$\checkmark$&$\checkmark$\\
$J^{\mathrm{(S)}13}_{ij}$&0	&0	&0	&0	&0	&0	&$\checkmark$&0	&0	&$\checkmark$&0	&0	&$\checkmark$&0	&$\checkmark$\\
$J^{\mathrm{(S)}14}_{ij}$&0	&0	&0	&0	&0	&$\checkmark$&0	&0	&$\checkmark$&0	&0	&$\checkmark$&0	&0	&$\checkmark$\\
$J^{\mathrm{(S)}15}_{ij}$&$\checkmark$&$\checkmark$&$\checkmark$&$\checkmark$&$\checkmark$&$\checkmark$&$\checkmark$&$\checkmark$&$\checkmark$&$\checkmark$&$\checkmark$&$\checkmark$&$\checkmark$&$\checkmark$&$\checkmark$\\
$J^{\mathrm{(S)}23}_{ij}$&0	&0	&0	&0	&0	&$\checkmark$&0	&0	&$\checkmark$&0	&0	&$\checkmark$&0	&0	&$\checkmark$\\
$J^{\mathrm{(S)}24}_{ij}$&0	&0	&0	&0	&0	&0	&$\checkmark$&0	&0	&$\checkmark$&0	&0	&$\checkmark$&0	&$\checkmark$\\
$J^{\mathrm{(S)}25}_{ij}$&0	&0	&0	&0	&0	&0	&0	&$\checkmark$&0	&0	&$\checkmark$&0	&0	&$\checkmark$&$\checkmark$\\
$J^{\mathrm{(S)}34}_{ij}$&0	&0	&0	&0	&0	&0	&0	&$\checkmark$&0	&0	&$\checkmark$&0	&0	&$\checkmark$&$\checkmark$\\
$J^{\mathrm{(S)}35}_{ij}$&0	&0	&0	&0	&0	&0	&$\checkmark$&0	&0	&$\checkmark$&0	&0	&$\checkmark$&0	&$\checkmark$\\
$J^{\mathrm{(S)}45}_{ij}$&0	&0	&0	&0	&0	&$\checkmark$&0	&0	&$\checkmark$&0	&0	&$\checkmark$&0	&0	&$\checkmark$	
\\ \hline
\# & 1 &  1 & 1 & 1 & 1 & 4 & 4 & 4 & 4 & 4 & 4 & 4 & 4 & 4 & 10
\\ \hline \hline 
\end{tabular}
}
\end{table*}

Next, we classify the quadrupole interactions according to the crystal symmetry. 
First, we show the classification of the antisymmetric interactions of the electric quadrupoles with the hexagonal bases for the $[100]$ bond under the point groups $mmm$, $2mm$, $m2m$, $mm2$, $222$, $2/m..$, $.2/m.$, $..2/m$, $2..$, $.2.$, $..2$, $m..$, $.m.$, $..m$, and $\bar{1}$ in Table.~\ref{tab:DM_Q}, where the 1st, 2nd, and 3rd axes in the point groups are $[001]$, $[100]$, and $[010]$, respectively. 
It is noted that the following classification results are also applied to the other three quadrupoles. 
Similarly to the DM interaction between the classical spins $(S^x_i, S^y_i, S^z_i)$, the antisymmetric quadrupole interactions vanish in the presence of the inversion symmetry $\bar{1}$.
In the noncentrosymmetric systems, however, they are symmetry allowed, and their components depend on the mirror and rotational symmetries~\cite{Kim_PhysRevB.87.205119, hosoi2020dzyaloshinskii}.
For example, the antisymmetric quadrupole interaction between the electric quadrupoles $(Q^2_i, Q^3_i)$, i.e., $J^{\mathrm{(AS)}32}_{ij}(Q^3_iQ^2_j-Q^2_iQ^3_j)$, has the same symmetry constraints in terms of the twofold rotations and mirrors as the DM interaction $D^{xy}_{ij}(S^x_iS^y_j-S^y_iS^x_j)$. 
This is because the electric quadrupoles $(Q^2_i, Q^3_i)$ and the spins (magnetic dipoles) $(S^x_i, S^y_i)$ are the bases of the two-dimensional irreducible representations $E^{+}_{1g}$ and $E^{-}_{1g}$ for the point group $6/mmm$, respectively; the superscript of the irreducible representation represents the time-reversal parity. 
Note that the different time-reversal symmetry of $E^{+}_{1g}$ and $E^{-}_{1g}$ does not lead to a difference in the bilinear interactions in Eq.~(\ref{eq:multiple_int}).
Similarly, the antisymmetric quadrupole interaction between $(Q^4_i, Q^5_i)$ also follows the same symmetry rules as $D^{xy}_{ij}(S^x_iS^y_j-S^y_iS^x_j)$, although $(Q^4_i, Q^5_i)$ belongs to the different irreducible representation $E^{+}_{2g}$. 
This is because the different sixfold rotational symmetry between $E_{1g}$ and $E_{2g}$ also leads to no difference in the bilinear interactions. 
The presence of the anisotropic quadrupole interaction indicates the possibility of the noncoplanar quadrupole orderings in analogy with the emergence of noncoplanar spin textures by the DM interaction~\cite{Akagi_PhysRevD.103.065008}.  
In addition, we find the emergent antisymmetric interactions between quadrupoles belonging to the different irreducible representations, such as $J^{{\rm (AS)}12}_{ij}$. 
Since various antisymmetric quadrupole interactions are possible, one can expect unconventional noncoplanar quadrupole textures that do not appear in the classical spin models. 
Furthermore, in contrast to noncoplanar spin textures, noncoplanar quadrupole textures are characterized by a different topological property, since the quadrupole space $(Q^1_i,Q^2_i,Q^3_i,Q^4_i,Q^5_i)$ is different from the classical spin space, the two-dimensional sphere $S^2$~\cite{mikushina2002dipole}.

Table~\ref{tab: Sym_Q} shows the classification of the symmetric quadrupole interaction. 
In contrast to the antisymmetric quadrupole interaction in Table~\ref{tab:DM_Q}, all the off-diagonal components in $J^{\mathrm{(S)}\alpha\beta}_{ij}$ are allowed in the presence of the inversion symmetry $\bar{1}$. 
Thus, the symmetric quadrupole interaction is present even in centrosymmetric lattice structures, which implies the appearance of multiple-$Q$ quadrupole orderings by the symmetric anisotropic interactions, as found in the case of the dipole interaction in Sec.~\ref{sec:int_dipole}~\cite{Hayami_doi:10.7566/JPSJ.89.103702, amoroso2020spontaneous, Hayami_PhysRevB.103.054422, yambe2021skyrmion, Hirschberger_10.1088/1367-2630/abdef9, Wang_PhysRevB.103.104408, Kato_PhysRevB.104.224405, amoroso2021tuning}.  
For $.2.$, $m..$, and $..m$ with the type I operations, the symmetry-allowed components in $J^{\mathrm{(S)}\alpha\beta}_{ij}$ are the same as those in $J^{\mathrm{(AS)}\alpha\beta}_{ij}$. 
Meanwhile, the symmetry-allowed components in $J^{\mathrm{(S)}\alpha\beta}_{ij}$ for $2..$, $..2$, and $.m.$ with the type II operations are different from those in $J^{\mathrm{(AS)}\alpha\beta}_{ij}$.

\subsection{Octupole Interactions}
\label{sec:int_octupole}

\begin{table*}
\centering
\caption{
\label{tab:hexagonal_O_AS}
Classification of the antisymmetric octupole interactions with the hexagonal bases for the $[100]$ bond.
The row ``\#" represents the number of the independent interaction parameters.
}
\scalebox{1.0}{
\begin{tabular}{cccccccccccccccc}
\hline \hline 
& $mmm$ & $2mm$ & $m2m$ & $mm2$ & $222$ & $2/m..$ & $.2/m.$ & $..2/m$ & $2..$ & $.2.$ & $..2$ & $m..$ & $.m.$ & $..m$ & $\bar{1}$   \\ \hline
$J^{\mathrm{(AS)}12}_{ij}$&0	&$\checkmark$&0	&0	&0	&0	&0	&0	&$\checkmark$&0	&0	&0	&$\checkmark$&$\checkmark$&0\\
$J^{\mathrm{(AS)}13}_{ij}$&0	&0	&0	&0	&$\checkmark$&0	&0	&0	&$\checkmark$&$\checkmark$&$\checkmark$&0	&0	&0	&0\\
$J^{\mathrm{(AS)}14}_{ij}$&0	&$\checkmark$&0	&0	&0	&0	&0	&0	&$\checkmark$&0	&0	&0	&$\checkmark$&$\checkmark$&0\\
$J^{\mathrm{(AS)}15}_{ij}$&0	&0	&0	&0	&$\checkmark$&0	&0	&0	&$\checkmark$&$\checkmark$&$\checkmark$&0	&0	&0	&0\\
$J^{\mathrm{(AS)}16}_{ij}$&0	&0	&0	&$\checkmark$&0	&0	&0	&0	&0	&0	&$\checkmark$&$\checkmark$&$\checkmark$&0	&0\\
$J^{\mathrm{(AS)}17}_{ij}$&0	&0	&$\checkmark$&0	&0	&0	&0	&0	&0	&$\checkmark$&0	&$\checkmark$&0	&$\checkmark$&0\\
$J^{\mathrm{(AS)}23}_{ij}$&0	&0	&0	&$\checkmark$&0	&0	&0	&0	&0	&0	&$\checkmark$&$\checkmark$&$\checkmark$&0	&0\\
$J^{\mathrm{(AS)}24}_{ij}$&0	&0	&$\checkmark$&0	&0	&0	&0	&0	&0	&$\checkmark$&0	&$\checkmark$&0	&$\checkmark$&0\\
$J^{\mathrm{(AS)}25}_{ij}$&0	&0	&0	&$\checkmark$&0	&0	&0	&0	&0	&0	&$\checkmark$&$\checkmark$&$\checkmark$&0	&0\\
$J^{\mathrm{(AS)}26}_{ij}$&0	&0	&0	&0	&$\checkmark$&0	&0	&0	&$\checkmark$&$\checkmark$&$\checkmark$&0	&0	&0	&0\\
$J^{\mathrm{(AS)}27}_{ij}$&0	&$\checkmark$&0	&0	&0	&0	&0	&0	&$\checkmark$&0	&0	&0	&$\checkmark$&$\checkmark$&0\\
$J^{\mathrm{(AS)}34}_{ij}$&0	&0	&0	&$\checkmark$&0	&0	&0	&0	&0	&0	&$\checkmark$&$\checkmark$&$\checkmark$&0	&0\\
$J^{\mathrm{(AS)}35}_{ij}$&0	&0	&$\checkmark$&0	&0	&0	&0	&0	&0	&$\checkmark$&0	&$\checkmark$&0	&$\checkmark$&0\\
$J^{\mathrm{(AS)}36}_{ij}$&0	&$\checkmark$&0	&0	&0	&0	&0	&0	&$\checkmark$&0	&0	&0	&$\checkmark$&$\checkmark$&0\\
$J^{\mathrm{(AS)}37}_{ij}$&0	&0	&0	&0	&$\checkmark$&0	&0	&0	&$\checkmark$&$\checkmark$&$\checkmark$&0	&0	&0	&0\\
$J^{\mathrm{(AS)}45}_{ij}$&0	&0	&0	&$\checkmark$&0	&0	&0	&0	&0	&0	&$\checkmark$&$\checkmark$&$\checkmark$&0	&0\\
$J^{\mathrm{(AS)}46}_{ij}$&0	&0	&0	&0	&$\checkmark$&0	&0	&0	&$\checkmark$&$\checkmark$&$\checkmark$&0	&0	&0	&0\\
$J^{\mathrm{(AS)}47}_{ij}$&0	&$\checkmark$&0	&0	&0	&0	&0	&0	&$\checkmark$&0	&0	&0	&$\checkmark$&$\checkmark$&0\\
$J^{\mathrm{(AS)}56}_{ij}$&0	&$\checkmark$&0	&0	&0	&0	&0	&0	&$\checkmark$&0	&0	&0	&$\checkmark$&$\checkmark$&0\\
$J^{\mathrm{(AS)}57}_{ij}$&0	&0	&0	&0	&$\checkmark$&0	&0	&0	&$\checkmark$&$\checkmark$&$\checkmark$&0	&0	&0	&0\\
$J^{\mathrm{(AS)}67}_{ij}$&0	&0	&0	&$\checkmark$&0	&0	&0	&0	&0	&0	&$\checkmark$&$\checkmark$&$\checkmark$ &0	&0
\\ \hline
\# & 0 &  6 & 3 & 6 & 6 & 0 & 0 & 0 & 12 & 9 & 12 & 9 & 12 & 9 & 0
\\ \hline \hline 
\end{tabular}
}
\end{table*}

\begin{table*}
\centering
\caption{
\label{tab:hexagonal_O_S}
Classification of the symmetric octupole interactions with the hexagonal bases for the $[100]$ bond.
The row ``\#" represents the number of the independent interaction parameters except for the diagonal components.
}
\scalebox{1.0}{
\begin{tabular}{cccccccccccccccc}
\hline \hline 
& $mmm$ & $2mm$ & $m2m$ & $mm2$ & $222$ & $2/m..$ & $.2/m.$ & $..2/m$ & $2..$ & $.2.$ & $..2$ & $m..$ & $.m.$ & $..m$ & $\bar{1}$   \\ \hline
$J^{\mathrm{(S)}12}_{ij}$&0	&0	&0	&0	&0	&0	&0	&$\checkmark$&0	&0	&$\checkmark$&0	&0	&$\checkmark$&$\checkmark$\\
$J^{\mathrm{(S)}13}_{ij}$&0	&0	&0	&0	&0	&0	&$\checkmark$&0	&0	&$\checkmark$&0	&0	&$\checkmark$&0	&$\checkmark$\\
$J^{\mathrm{(S)}14}_{ij}$&0	&0	&0	&0	&0	&0	&0	&$\checkmark$&0	&0	&$\checkmark$&0	&0	&$\checkmark$&$\checkmark$\\
$J^{\mathrm{(S)}15}_{ij}$&0	&0	&0	&0	&0	&0	&$\checkmark$&0	&0	&$\checkmark$&0	&0	&$\checkmark$&0	&$\checkmark$\\
$J^{\mathrm{(S)}16}_{ij}$&0	&0	&0	&0	&0	&$\checkmark$&0	&0	&$\checkmark$&0	&0	&$\checkmark$&0	&0	&$\checkmark$\\
$J^{\mathrm{(S)}17}_{ij}$&$\checkmark$&$\checkmark$&$\checkmark$&$\checkmark$&$\checkmark$&$\checkmark$&$\checkmark$&$\checkmark$&$\checkmark$&$\checkmark$&$\checkmark$&$\checkmark$&$\checkmark$&$\checkmark$&$\checkmark$\\
$J^{\mathrm{(S)}23}_{ij}$&0	&0	&0	&0	&0	&$\checkmark$&0	&0	&$\checkmark$&0	&0	&$\checkmark$&0	&0	&$\checkmark$\\
$J^{\mathrm{(S)}24}_{ij}$&$\checkmark$&$\checkmark$&$\checkmark$&$\checkmark$&$\checkmark$&$\checkmark$&$\checkmark$&$\checkmark$&$\checkmark$&$\checkmark$&$\checkmark$&$\checkmark$&$\checkmark$&$\checkmark$&$\checkmark$\\
$J^{\mathrm{(S)}25}_{ij}$&0	&0	&0	&0	&0	&$\checkmark$&0	&0	&$\checkmark$&0	&0	&$\checkmark$&0	&0	&$\checkmark$\\
$J^{\mathrm{(S)}26}_{ij}$&0	&0	&0	&0	&0	&0	&$\checkmark$&0	&0	&$\checkmark$&0	&0	&$\checkmark$&0	&$\checkmark$\\
$J^{\mathrm{(S)}27}_{ij}$&0	&0	&0	&0	&0	&0	&0	&$\checkmark$&0	&0	&$\checkmark$&0	&0	&$\checkmark$&$\checkmark$\\
$J^{\mathrm{(S)}34}_{ij}$&0	&0	&0	&0	&0	&$\checkmark$&0	&0	&$\checkmark$&0	&0	&$\checkmark$&0	&0	&$\checkmark$\\
$J^{\mathrm{(S)}35}_{ij}$&$\checkmark$&$\checkmark$&$\checkmark$&$\checkmark$&$\checkmark$&$\checkmark$&$\checkmark$&$\checkmark$&$\checkmark$&$\checkmark$&$\checkmark$&$\checkmark$&$\checkmark$&$\checkmark$&$\checkmark$\\
$J^{\mathrm{(S)}36}_{ij}$&0	&0	&0	&0	&0	&0	&0	&$\checkmark$&0	&0	&$\checkmark$&0	&0	&$\checkmark$&$\checkmark$\\
$J^{\mathrm{(S)}37}_{ij}$&0	&0	&0	&0	&0	&0	&$\checkmark$&0	&0	&$\checkmark$&0	&0	&$\checkmark$&0	&$\checkmark$\\
$J^{\mathrm{(S)}45}_{ij}$&0	&0	&0	&0	&0	&$\checkmark$&0	&0	&$\checkmark$&0	&0	&$\checkmark$&0	&0	&$\checkmark$\\
$J^{\mathrm{(S)}46}_{ij}$&0	&0	&0	&0	&0	&0	&$\checkmark$&0	&0	&$\checkmark$&0	&0	&$\checkmark$&0	&$\checkmark$\\
$J^{\mathrm{(S)}47}_{ij}$&0	&0	&0	&0	&0	&0	&0	&$\checkmark$&0	&0	&$\checkmark$&0	&0	&$\checkmark$&$\checkmark$\\
$J^{\mathrm{(S)}56}_{ij}$&0	&0	&0	&0	&0	&0	&0	&$\checkmark$&0	&0	&$\checkmark$&0	&0	&$\checkmark$&$\checkmark$\\
$J^{\mathrm{(S)}57}_{ij}$&0	&0	&0	&0	&0	&0	&$\checkmark$&0	&0	&$\checkmark$&0	&0	&$\checkmark$&0	&$\checkmark$\\
$J^{\mathrm{(S)}67}_{ij}$&0	&0	&0	&0	&0	&$\checkmark$&0	&0	&$\checkmark$&0	&0	&$\checkmark$&0	&0	&$\checkmark$
\\ \hline
\# & 3 &  3 & 3 & 3 & 3 & 9 & 9 & 9 & 9 & 9 & 9 & 9 & 9 & 9 & 21
\\ \hline \hline 
\end{tabular}
}
\end{table*}

The classification of the antisymmetric and symmetric octupole interactions is shown in Tables~\ref{tab:hexagonal_O_AS} and \ref{tab:hexagonal_O_S}, respectively. 
Similarly to the classification of the dipole and quadrupole interactions, the antisymmetric (symmetric) octupole interaction is prohibited (allowed) in the presence of the inversion symmetry $\bar{1}$. 
For the other symmetry operations, the tendency to have nonzero components of the antisymmetric and symmetric octupole interactions is similar to the dipole and quadrupole interactions.

\subsection{Hexadecapole Interactions}
\label{sec:int_hexadecapole}

\begin{table*}
\centering
\caption{
\label{tab:hexagonal_H_AS}
Classification of the antisymmetric hexadecapole interactions with the hexagonal bases for the $[100]$ bond.
The row ``\#" represents the number of the independent interaction parameters.
}
\scalebox{1.0}{
\begin{tabular}{cccccccccccccccc}
\hline \hline 
& $mmm$ & $2mm$ & $m2m$ & $mm2$ & $222$ & $2/m..$ & $.2/m.$ & $..2/m$ & $2..$ & $.2.$ & $..2$ & $m..$ & $.m.$ & $..m$ & $\bar{1}$   \\ \hline
$J^{\mathrm{(AS)}12}_{ij}$&0	&0	&0	&0	&$\checkmark$&0	&0	&0	&$\checkmark$&$\checkmark$&$\checkmark$&0	&0	&0	&0\\
$J^{\mathrm{(AS)}13}_{ij}$&0	&$\checkmark$&0	&0	&0	&0	&0	&0	&$\checkmark$&0	&0	&0	&$\checkmark$&$\checkmark$&0\\
$J^{\mathrm{(AS)}14}_{ij}$&0	&$\checkmark$&0	&0	&0	&0	&0	&0	&$\checkmark$&0	&0	&0	&$\checkmark$&$\checkmark$&0\\
$J^{\mathrm{(AS)}15}_{ij}$&0	&0	&0	&0	&$\checkmark$&0	&0	&0	&$\checkmark$&$\checkmark$&$\checkmark$&0	&0	&0	&0\\
$J^{\mathrm{(AS)}16}_{ij}$&0	&0	&$\checkmark$&0	&0	&0	&0	&0	&0	&$\checkmark$&0	&$\checkmark$&0	&$\checkmark$&0\\
$J^{\mathrm{(AS)}17}_{ij}$&0	&0	&0	&$\checkmark$&0	&0	&0	&0	&0	&0	&$\checkmark$&$\checkmark$&$\checkmark$&0	&0\\
$J^{\mathrm{(AS)}18}_{ij}$&0	&0	&0	&$\checkmark$&0	&0	&0	&0	&0	&0	&$\checkmark$&$\checkmark$&$\checkmark$&0	&0\\
$J^{\mathrm{(AS)}19}_{ij}$&0	&0	&$\checkmark$&0	&0	&0	&0	&0	&0	&$\checkmark$&0	&$\checkmark$&0	&$\checkmark$&0\\
$J^{\mathrm{(AS)}23}_{ij}$&0	&0	&0	&$\checkmark$&0	&0	&0	&0	&0	&0	&$\checkmark$&$\checkmark$&$\checkmark$&0	&0\\
$J^{\mathrm{(AS)}24}_{ij}$&0	&0	&0	&$\checkmark$&0	&0	&0	&0	&0	&0	&$\checkmark$&$\checkmark$&$\checkmark$&0	&0\\
$J^{\mathrm{(AS)}25}_{ij}$&0	&0	&$\checkmark$&0	&0	&0	&0	&0	&0	&$\checkmark$&0	&$\checkmark$&0	&$\checkmark$&0\\
$J^{\mathrm{(AS)}26}_{ij}$&0	&0	&0	&0	&$\checkmark$&0	&0	&0	&$\checkmark$&$\checkmark$&$\checkmark$&0	&0	&0	&0\\
$J^{\mathrm{(AS)}27}_{ij}$&0	&$\checkmark$&0	&0	&0	&0	&0	&0	&$\checkmark$&0	&0	&0	&$\checkmark$&$\checkmark$&0\\
$J^{\mathrm{(AS)}28}_{ij}$&0	&$\checkmark$&0	&0	&0	&0	&0	&0	&$\checkmark$&0	&0	&0	&$\checkmark$&$\checkmark$&0\\
$J^{\mathrm{(AS)}29}_{ij}$&0	&0	&0	&0	&$\checkmark$&0	&0	&0	&$\checkmark$&$\checkmark$&$\checkmark$&0	&0	&0	&0\\
$J^{\mathrm{(AS)}34}_{ij}$&0	&0	&$\checkmark$&0	&0	&0	&0	&0	&0	&$\checkmark$&0	&$\checkmark$&0	&$\checkmark$&0\\
$J^{\mathrm{(AS)}35}_{ij}$&0	&0	&0	&$\checkmark$&0	&0	&0	&0	&0	&0	&$\checkmark$&$\checkmark$&$\checkmark$&0	&0\\
$J^{\mathrm{(AS)}36}_{ij}$&0	&$\checkmark$&0	&0	&0	&0	&0	&0	&$\checkmark$&0	&0	&0	&$\checkmark$&$\checkmark$&0\\
$J^{\mathrm{(AS)}37}_{ij}$&0	&0	&0	&0	&$\checkmark$&0	&0	&0	&$\checkmark$&$\checkmark$&$\checkmark$&0	&0	&0	&0\\
$J^{\mathrm{(AS)}38}_{ij}$&0	&0	&0	&0	&$\checkmark$&0	&0	&0	&$\checkmark$&$\checkmark$&$\checkmark$&0	&0	&0	&0\\
$J^{\mathrm{(AS)}39}_{ij}$	&0 &$\checkmark$&0	&0	&0	&0	&0	&0	&$\checkmark$&0	&0	&0	&$\checkmark$&$\checkmark$&0\\
$J^{\mathrm{(AS)}45}_{ij}$&0	&0	&0	&$\checkmark$&0	&0	&0	&0	&0	&0	&$\checkmark$&$\checkmark$&$\checkmark$&0	&0\\
$J^{\mathrm{(AS)}46}_{ij}$&0	&$\checkmark$&0	&0	&0	&0	&0	&0	&$\checkmark$&0	&0	&0	&$\checkmark$&$\checkmark$&0\\
$J^{\mathrm{(AS)}47}_{ij}$&0	&0	&0	&0	&$\checkmark$&0	&0	&0	&$\checkmark$&$\checkmark$&$\checkmark$&0	&0	&0	&0\\
$J^{\mathrm{(AS)}48}_{ij}$&0	&0	&0	&0	&$\checkmark$&0	&0	&0	&$\checkmark$&$\checkmark$&$\checkmark$&0	&0	&0	&0\\
$J^{\mathrm{(AS)}49}_{ij}$&0	&$\checkmark$&0	&0	&0	&0	&0	&0	&$\checkmark$&0	&0	&0	&$\checkmark$&$\checkmark$&0\\
$J^{\mathrm{(AS)}56}_{ij}$&0	&0	&0	&0	&$\checkmark$&0	&0	&0	&$\checkmark$&$\checkmark$&$\checkmark$&0	&0	&0	&0\\
$J^{\mathrm{(AS)}57}_{ij}$&0	&$\checkmark$&0	&0	&0	&0	&0	&0	&$\checkmark$&0	&0	&0	&$\checkmark$&$\checkmark$&0\\
$J^{\mathrm{(AS)}58}_{ij}$&0	&$\checkmark$&0	&0	&0	&0	&0	&0	&$\checkmark$&0	&0	&0	&$\checkmark$&$\checkmark$&0\\
$J^{\mathrm{(AS)}59}_{ij}$&0	&0	&0	&0	&$\checkmark$&0	&0	&0	&$\checkmark$&$\checkmark$&$\checkmark$&0	&0	&0	&0\\
$J^{\mathrm{(AS)}67}_{ij}$&0	&0	&0	&$\checkmark$&0	&0	&0	&0	&0	&0	&$\checkmark$&$\checkmark$&$\checkmark$&0	&0\\
$J^{\mathrm{(AS)}68}_{ij}$&0	&0	&0	&$\checkmark$&0	&0	&0	&0	&0	&0	&$\checkmark$&$\checkmark$&$\checkmark$&0	&0\\
$J^{\mathrm{(AS)}69}_{ij}$&0	&0	&$\checkmark$&0	&0	&0	&0	&0	&0	&$\checkmark$&0	&$\checkmark$&0	&$\checkmark$&0\\
$J^{\mathrm{(AS)}78}_{ij}$&0	&0	&$\checkmark$&0	&0	&0	&0	&0	&0	&$\checkmark$&0	&$\checkmark$&0	&$\checkmark$&0\\
$J^{\mathrm{(AS)}79}_{ij}$&0	&0	&0	&$\checkmark$&0	&0	&0	&0	&0	&0	&$\checkmark$&$\checkmark$&$\checkmark$&0	&0\\
$J^{\mathrm{(AS)}89}_{ij}$&0	&0	&0	&$\checkmark$&0	&0	&0	&0	&0	&0	&$\checkmark$&$\checkmark$&$\checkmark$&0	&0
\\ \hline
\# & 0 &  10 & 6 & 10 & 10 & 0 & 0 & 0 & 20 & 16 & 20 & 16 & 20 & 16 & 0
\\ \hline \hline 
\end{tabular}
}
\end{table*}

\begin{table*}
\centering
\caption{
\label{tab:hexagonal_H_S}
Classification of the symmetric hexadecapole interactions with the hexagonal bases for the $[100]$ bond.
The row ``\#" represents the number of the independent interaction parameters except for the diagonal components.
}
\scalebox{0.9}{
\begin{tabular}{cccccccccccccccc}
\hline \hline 
& $mmm$ & $2mm$ & $m2m$ & $mm2$ & $222$ & $2/m..$ & $.2/m.$ & $..2/m$ & $2..$ & $.2.$ & $..2$ & $m..$ & $.m.$ & $..m$ & $\bar{1}$   \\ \hline
$J^{\mathrm{(S)}12}_{ij}$&0	&0	&0	&0	&0	&0	&$\checkmark$&0	&0	&$\checkmark$&0	&0	&$\checkmark$&0	&$\checkmark$\\
$J^{\mathrm{(S)}13}_{ij}$&0	&0	&0	&0	&0	&0	&0	&$\checkmark$&0	&0	&$\checkmark$&0	&0	&$\checkmark$&$\checkmark$\\
$J^{\mathrm{(S)}14}_{ij}$&0	&0	&0	&0	&0	&0	&0	&$\checkmark$&0	&0	&$\checkmark$&0	&0	&$\checkmark$&$\checkmark$\\
$J^{\mathrm{(S)}15}_{ij}$&0	&0	&0	&0	&0	&0	&$\checkmark$&0	&0	&$\checkmark$&0	&0	&$\checkmark$&0	&$\checkmark$\\
$J^{\mathrm{(S)}16}_{ij}$&$\checkmark$&$\checkmark$&$\checkmark$&$\checkmark$&$\checkmark$&$\checkmark$&$\checkmark$&$\checkmark$&$\checkmark$&$\checkmark$&$\checkmark$&$\checkmark$&$\checkmark$&$\checkmark$&$\checkmark$\\
$J^{\mathrm{(S)}17}_{ij}$&0	&0	&0	&0	&0	&$\checkmark$&0	&0	&$\checkmark$&0	&0	&$\checkmark$&0	&0	&$\checkmark$\\
$J^{\mathrm{(S)}18}_{ij}$&0	&0	&0	&0	&0	&$\checkmark$&0	&0	&$\checkmark$&0	&0	&$\checkmark$&0	&0	&$\checkmark$\\
$J^{\mathrm{(S)}19}_{ij}$&$\checkmark$&$\checkmark$&$\checkmark$&$\checkmark$&$\checkmark$&$\checkmark$&$\checkmark$&$\checkmark$&$\checkmark$&$\checkmark$&$\checkmark$&$\checkmark$&$\checkmark$&$\checkmark$&$\checkmark$\\
$J^{\mathrm{(S)}23}_{ij}$&0	&0	&0	&0	&0	&$\checkmark$&0	&0	&$\checkmark$&0	&0	&$\checkmark$&0	&0	&$\checkmark$\\
$J^{\mathrm{(S)}24}_{ij}$&0	&0	&0	&0	&0	&$\checkmark$&0	&0	&$\checkmark$&0	&0	&$\checkmark$&0	&0	&$\checkmark$\\
$J^{\mathrm{(S)}25}_{ij}$&$\checkmark$&$\checkmark$&$\checkmark$&$\checkmark$&$\checkmark$&$\checkmark$&$\checkmark$&$\checkmark$&$\checkmark$&$\checkmark$&$\checkmark$&$\checkmark$&$\checkmark$&$\checkmark$&$\checkmark$\\
$J^{\mathrm{(S)}26}_{ij}$&0	&0	&0	&0	&0	&0	&$\checkmark$&0	&0	&$\checkmark$&0	&0	&$\checkmark$&0	&$\checkmark$\\
$J^{\mathrm{(S)}27}_{ij}$&0	&0	&0	&0	&0	&0	&0	&$\checkmark$&0	&0	&$\checkmark$&0	&0	&$\checkmark$&$\checkmark$\\
$J^{\mathrm{(S)}28}_{ij}$&0	&0	&0	&0	&0	&0	&0	&$\checkmark$&0	&0	&$\checkmark$&0	&0	&$\checkmark$&$\checkmark$\\
$J^{\mathrm{(S)}29}_{ij}$&0	&0	&0	&0	&0	&0	&$\checkmark$&0	&0	&$\checkmark$&0	&0	&$\checkmark$&0	&$\checkmark$\\
$J^{\mathrm{(S)}34}_{ij}$&$\checkmark$&$\checkmark$&$\checkmark$&$\checkmark$&$\checkmark$&$\checkmark$&$\checkmark$&$\checkmark$&$\checkmark$&$\checkmark$&$\checkmark$&$\checkmark$&$\checkmark$&$\checkmark$&$\checkmark$\\
$J^{\mathrm{(S)}35}_{ij}$&0	&0	&0	&0	&0	&$\checkmark$&0	&0	&$\checkmark$&0	&0	&$\checkmark$&0	&0	&$\checkmark$\\
$J^{\mathrm{(S)}36}_{ij}$&0	&0	&0	&0	&0	&0	&0	&$\checkmark$&0	&0	&$\checkmark$&0	&0	&$\checkmark$&$\checkmark$\\
$J^{\mathrm{(S)}37}_{ij}$&0	&0	&0	&0	&0	&0	&$\checkmark$&0	&0	&$\checkmark$&0	&0	&$\checkmark$&0	&$\checkmark$\\
$J^{\mathrm{(S)}38}_{ij}$&0	&0	&0	&0	&0	&0	&$\checkmark$&0	&0	&$\checkmark$&0	&0	&$\checkmark$&0	&$\checkmark$\\
$J^{\mathrm{(S)}39}_{ij}$&0	&0	&0	&0	&0	&0	&0	&$\checkmark$&0	&0	&$\checkmark$&0	&0	&$\checkmark$&$\checkmark$\\
$J^{\mathrm{(S)}45}_{ij}$&0	&0	&0	&0	&0	&$\checkmark$&0	&0	&$\checkmark$&0	&0	&$\checkmark$&0	&0	&$\checkmark$\\
$J^{\mathrm{(S)}46}_{ij}$&0	&0	&0	&0	&0	&0	&0	&$\checkmark$&0	&0	&$\checkmark$&0	&0	&$\checkmark$&$\checkmark$\\
$J^{\mathrm{(S)}47}_{ij}$&0	&0	&0	&0	&0	&0	&$\checkmark$&0	&0	&$\checkmark$&0	&0	&$\checkmark$&0	&$\checkmark$\\
$J^{\mathrm{(S)}48}_{ij}$&0	&0	&0	&0	&0	&0	&$\checkmark$&0	&0	&$\checkmark$&0	&0	&$\checkmark$&0	&$\checkmark$\\
$J^{\mathrm{(S)}49}_{ij}$&0	&0	&0	&0	&0	&0	&0	&$\checkmark$&0	&0	&$\checkmark$&0	&0	&$\checkmark$&$\checkmark$\\
$J^{\mathrm{(S)}56}_{ij}$&0	&0	&0	&0	&0	&0	&$\checkmark$&0	&0	&$\checkmark$&0	&0	&$\checkmark$&0	&$\checkmark$\\
$J^{\mathrm{(S)}57}_{ij}$&0	&0	&0	&0	&0	&0	&0	&$\checkmark$&0	&0	&$\checkmark$&0	&0	&$\checkmark$&$\checkmark$\\
$J^{\mathrm{(S)}58}_{ij}$&0	&0	&0	&0	&0	&0	&0	&$\checkmark$&0	&0	&$\checkmark$&0	&0	&$\checkmark$&$\checkmark$\\
$J^{\mathrm{(S)}59}_{ij}$&0	&0	&0	&0	&0	&0	&$\checkmark$&0	&0	&$\checkmark$&0	&0	&$\checkmark$&0	&$\checkmark$\\
$J^{\mathrm{(S)}67}_{ij}$&0	&0	&0	&0	&0	&$\checkmark$&0	&0	&$\checkmark$&0	&0	&$\checkmark$&0	&0	&$\checkmark$\\
$J^{\mathrm{(S)}68}_{ij}$&0	&0	&0	&0	&0	&$\checkmark$&0	&0	&$\checkmark$&0	&0	&$\checkmark$&0	&0	&$\checkmark$\\
$J^{\mathrm{(S)}69}_{ij}$&$\checkmark$&$\checkmark$&$\checkmark$&$\checkmark$&$\checkmark$&$\checkmark$&$\checkmark$&$\checkmark$&$\checkmark$&$\checkmark$&$\checkmark$&$\checkmark$&$\checkmark$&$\checkmark$&$\checkmark$\\
$J^{\mathrm{(S)}78}_{ij}$&$\checkmark$&$\checkmark$&$\checkmark$&$\checkmark$&$\checkmark$&$\checkmark$&$\checkmark$&$\checkmark$&$\checkmark$&$\checkmark$&$\checkmark$&$\checkmark$&$\checkmark$&$\checkmark$&$\checkmark$\\
$J^{\mathrm{(S)}79}_{ij}$&0	&0	&0	&0	&0	&$\checkmark$&0	&0	&$\checkmark$&0	&0	&$\checkmark$&0	&0	&$\checkmark$\\
$J^{\mathrm{(S)}89}_{ij}$&0	&0	&0	&0	&0	&$\checkmark$&0	&0	&$\checkmark$&0	&0	&$\checkmark$&0	&0	&$\checkmark$
\\ \hline
\# & 6 &  6 & 6 & 6 & 6 & 16 & 16 & 16 & 16 & 16 & 16 & 16 & 16 & 16 & 36
\\ \hline \hline 
\end{tabular}
}
\end{table*}

Finally, we present the classification results of the antisymmetric and symmetric hexadecapole interactions in Tables~\ref{tab:hexagonal_H_AS} and \ref{tab:hexagonal_H_S}, respectively. 
The tendency of the symmetry-allowed components in $J^{\mathrm{(S)}\alpha\beta}_{ij}$ and $J^{\mathrm{(S)}\alpha\beta}_{ij}$ resembles the other multipole cases.

\section{Application to Triple-$Q$ Quadrupole State}
\label{sec:Class_2}

Based on the symmetry analysis in Sec.~\ref{sec:int}, we construct an effective multipole model for a given symmetry and lattice structure, and demonstrate that the symmetry-allowed anisotropic multipole interaction becomes the origin of multiple-$Q$ multipole orderings. 
We consider the two-dimensional triangular-lattice system under the $C_{\rm 6v}$ symmetry, where the lattice constant of the triangular lattice is set to unity; the primitive translational vectors are given by $\bm{a}_1=(1,0)$ and $\bm{a}_2=(-1/2, \sqrt{3}/2)$. 
We adopt the spin-1 system, where the magnetic dipole and electric quadrupole moments are activated at each lattice site; $Q^1$, $(Q^2, Q^3)$, and $(Q^4,Q^5)$ belong to the irreducible representations $A_1$, $E_1$, and $E_2$ under the $C_{\rm 6v}$ symmetry, respectively. 

The spin-1 model with the nearest-neighbor interaction is given by~\cite{Fath_PhysRevB.51.3620, Schollwock_PhysRevB.53.3304, Harada_PhysRevB.65.052403, Lauchli_PhysRevLett.97.087205, tsunetsugu2006spin, tsunetsugu2007spin, Smerald_PhysRevB.88.184430, Pohle_PhysRevB.107.L140403} 
\begin{align}
\mathcal{H} = \mathcal{H}_{\rm BBQ} + \mathcal{H}_{\rm DM}, 
\end{align}
where $\mathcal{H}_{\rm BBQ}$ and $\mathcal{H}_{\rm DM}$ stand for the bilinear-biquadratic Hamiltonian and DM-type antisymmetric quadrupole Hamiltonian, respectively. 
The former $\mathcal{H}_{\rm BBQ}$ is given by~\cite{Remund_PhysRevResearch.4.033106} 
\begin{align}
\label{eq: Ham_BBQ}
\mathcal{H}_{\rm BBQ} = \sum_{\langle i,j \rangle} \left[ 
J_1 \bm{S}_i\cdot \bm{S}_j + J_2 (\bm{S}_i\cdot \bm{S}_j)^2 
\right],
\end{align}
where $\bm{S}_i$ rerpesents the spin-1 moment at site $i$; $J_1$ and $J_2$ correspond to the bilinear and biquadratic interactions, respectively. 
The model in Eq.~(\ref{eq: Ham_BBQ}) is rewritten by using the quadrupole moments $Q^1_i$--$Q^5_i$ as 
\begin{align}
\mathcal{H}_{\rm BBQ}= \sum_{\langle i,j \rangle} \left(J_1 -\frac{J_2}{2} \right) \bm{S}_i\cdot \bm{S}_j 
+ \frac{J_2}{2}\bm{Q}_i\cdot \bm{Q}_j + \frac{4J_2}{3}, 
\end{align}
where the quadrupole moment $\bm{Q}_i =(Q^1_i, Q^2_i, Q^3_i, Q^4_i, Q^5_i)$ is defined as 
\begin{align}
Q^1_i &= \frac{1}{2\sqrt{3}}(2Q^{zz}_i-Q^{xx}_i-Q^{yy}_i) \nonumber \\
       &= \frac{1}{\sqrt{3}}[2 (S^z_i)^2-(S^x_i)^2- (S^y_i)^2], \\
Q^2_i &= Q^{zx}_i= S^z_i S^x_i + S^x_i S^z_i, \\
Q^3_i &= Q^{yz}_i= S^y_i S^z_i + S^z_i S^y_i,\\
Q^4_i &= Q^{xy}_i= S^x_i S^y_i + S^y_i S^x_i, \\
Q^5_i &= \frac{1}{2} (Q^{xx}_i-Q^{yy}_i)= (S^x_i)^2- (S^y_i)^2. 
\end{align}
In order to focus on the multiple-$Q$ quadrupole ordering, we neglect the dipole interaction by setting $J_2 = 2J_1$. 

We add the modulation by the antisymmetric quadrupole interaction $\mathcal{H}_{\rm DM}$, which is symmetry-allowed under $C_{\rm 6v}$. 
Since the symmetry of the [100] bond is characterized by $2mm$, nonzero components in $J^{\mathrm{(AS)}\alpha \beta}_{ii+\boldsymbol{a}_1}$ are given by $J^{\mathrm{(AS)}12}_{ii+\boldsymbol{a}_1}=D^{12}$, $J^{\mathrm{(AS)}52}_{ii+\boldsymbol{a}_1}=D^{52}$, and $J^{\mathrm{(AS)}34}_{ii+\boldsymbol{a}_1}=D^{34}$, as shown in Table~\ref{tab:DM_Q}. 
Specifically, $\mathcal{H}_{\rm DM}$ for the [100] bond is given by 
\begin{align}
\mathcal{H}^{[100]}_{\rm DM} = &D^{12}(Q^1_i Q^2_{i+\boldsymbol{a}_1} - Q^2_i Q^1_{i+\boldsymbol{a}_1})\nonumber \\ 
&+ D^{52}(Q^5_i Q^2_{i+\boldsymbol{a}_1} - Q^2_i Q^5_{i+\boldsymbol{a}_1}) \nonumber \\ 
&+ D^{43}(Q^4_i Q^3_{i+\boldsymbol{a}_1} - Q^3_i Q^4_{i+\boldsymbol{a}_1}), 
\end{align}
where $\mathcal{H}_{\rm DM}$ for the other bond directions is obtained by imposing the sixfold rotational symmetry around the $i$th lattice site.

To efficiently evaluate the quadrupole moment in the spin-1 system, we introduce time-reversal invariant basis $\ket{\bm{d}}$ as~\cite{Remund_PhysRevResearch.4.033106} 
\begin{align}
\ket{\bm{d}} = \sum_{\alpha=x,y,z}d^*_\alpha \ket{\alpha},
\end{align}
where 
\begin{align}
\ket{x} = \frac{i}{\sqrt{2}}(\ket{1}-\ket{\bar{1}}), 
\ket{y} = \frac{1}{\sqrt{2}}(\ket{1}+\ket{\bar{1}}), 
\ket{z} = -i \ket{0}, 
\end{align}
and $d^*_\alpha$ are complex coefficients to satisfy $\bm{d}^* \cdot \bm{d} = 1$; $d^*_\alpha$ is refereed to as a director. 
$d^*_\alpha$ is separated into real and imagiary parts as $\bm{d}^* = \bm{u} + i \bm{v}$. 

When the dipole moments can be neglected, the director $\bm{d}$ is represented by either purely real or imaginary; we here set $\bm{v}=0$. 
Then, the quadrupole moment is expressed by using $\bm{u}$, which is given by 
\begin{align}
Q^1_i &= \frac{1}{\sqrt{3}}[-2 (u^z_i)^2+(u^x_i)^2+ (u^y_i)^2], \\
Q^2_i &= -2u^z_i u^x_i, \\
Q^3_i &= -2u^y_i u^z_i,\\
Q^4_i &= -2u^x_i u^y_i, \\
Q^5_i &= -(u^x_i)^2+ (u^y_i)^2. 
\end{align}
Then, the Hamiltonian for the [100] bond is rewritten as 
\begin{align}
\mathcal{H}^{[100]} &= - 2J (\bm{u}_i \cdot \bm{u}_{i+\boldsymbol{a}_1})^2 \nonumber \\
&-4 D(\bm{u}_i \cdot \bm{u}_{i+\boldsymbol{a}_1})(\bm{u}_i \times \bm{u}_{i+\boldsymbol{a}_1})^y + \frac{2J}{3},
\end{align}
where we set $D^{12}=-\sqrt{3}D$ and $D^{52}=D^{43}=D$ in order to simplify the DM Hamiltonian. 
Thus, the model includes the isotropic and polar biquadratic interactions in terms of $\bm{u}_i$~\cite{brinker2019chiral, Hayami_PhysRevB.105.024413}. 

To obtain the lowest-energy spin configuration of the above quadrupole model, we perform the Monte Carlo simulations. 
In the simulations, we set $u^x = \sqrt{1-\theta^2}\cos \phi$, $u^y = \sqrt{1-\theta^2}\sin \phi$, and $u^z= \theta$ for $0 \leq \theta \leq 1$ and $0 \leq \phi < 2\pi$, and update $\theta$ and $\phi$ in each site based on the stadanrd Metropolis algorithm; it is noted that $\bm{u}$ is characterized by two valuables as found in the classical Heisenberg spin. 
Then, we calculate the director $\bm{u}$, and evaluate the energy. 
In order to avoid the system becoming trapped in a metastable state, we gradually decrease the temperature from a high temperature $T= 2$ to a low temperature $T=10^{-4} $, following the manner based on the simulated annealing. 
For the obtained quadrupole configurations, we calculate the director and quadrupole structure factors, which are given by 
\begin{align}
S_O^{\alpha} (\bm{q}) &= \langle O^\alpha_{\bm{q}}O^\alpha_{-\bm{q}} \rangle, \\
O^\alpha_{\bm{q}}&= \frac{1}{\sqrt{N}}\sum_i O_i^\alpha e^{i\bm{q} \cdot \bm{r}_i}, 
\end{align}
where $O=u$ for the director and $O=Q$ for the quadrupole and $N$ is the system size. 
We set $N=24^2$ under the periodic boundary conditions.

\begin{figure}[tb!]
\begin{center}
\includegraphics[width=1.0\hsize]{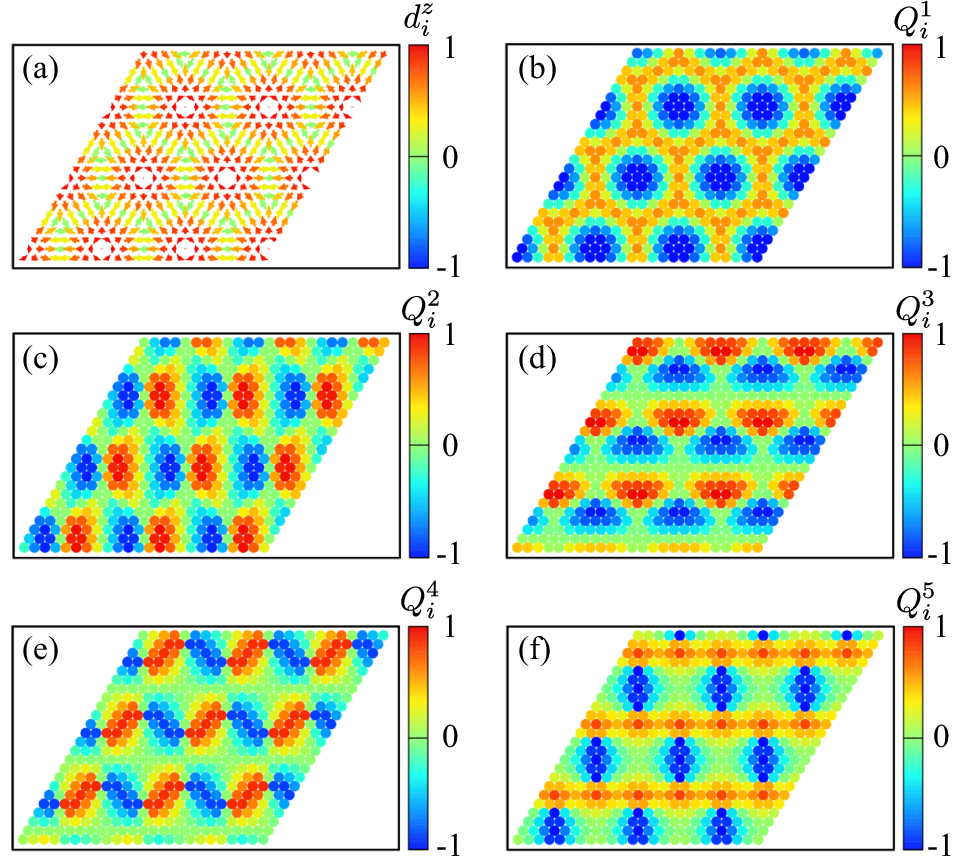} 
\caption{
\label{fig: real} 
(Color online) Real-space configurations of (a) the director $\bm{d}$, (b) $Q^1$, (c) $Q^2$, (d) $Q^3$, (e) $Q^4$, and (f) $Q^5$. 
In (a), the arrows represent the direction of the director vector and the color represents the $z$ component. 
}
\end{center}
\end{figure}

\begin{figure}[tb!]
\begin{center}
\includegraphics[width=0.7\hsize]{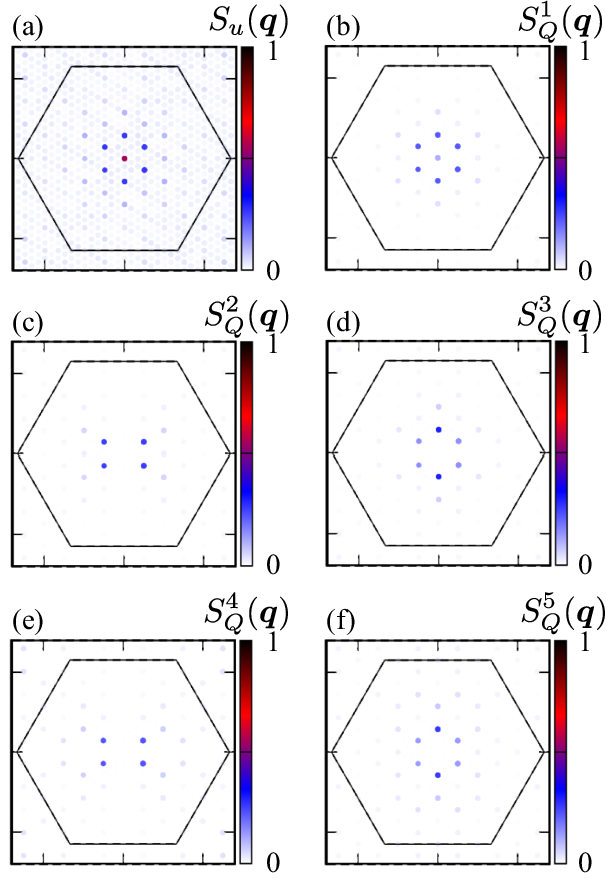} 
\caption{
\label{fig: qspace} 
(Color online) The structure factors of (a) the director, (b) $Q^1$, (c) $Q^2$, (d) $Q^3$, (e) $Q^4$, and (f) $Q^5$. 
}
\end{center}
\end{figure}

In the calculations, we set $J=1$ and $D=2/(1+\sqrt{3})\simeq 0.732$. 
Figure~\ref{fig: real} shows the real-space configurations of the director and the quadrupoles $(Q^1, Q^2, Q^3, Q^4, Q^5)$, which are obtained by the simulated annealing. 
As shown in Fig.~\ref{fig: real}(a), we obtain the single-domain state consisting of the periodic alignment of the vortex in terms of the director; the sixfold symmetric structure indicates the emergence of the triple-$Q$ structure. 
Indeed, the triple-$Q$ peaks at $\bm{Q}_1=[0, \pi/(2\sqrt{3})]$, $\bm{Q}_2=[-\sqrt{3}\pi/(4\sqrt{3}), -\pi/(4\sqrt{3})]$, and $\bm{Q}_3=[\sqrt{3}\pi/(4\sqrt{3}), -\pi/(4\sqrt{3})]$ are found in the director structure factor, as shown in Fig.~\ref{fig: qspace}(a), where we sum over each component when calculating the director structure factor. 
Accordingly, each quadrupole component also shows multiple-$Q$ peaks, as shown in Figs.~\ref{fig: qspace}(b)--\ref{fig: qspace}(f); the $Q^1$ component shows the threefold-symmetric peaks, while the $Q^2$--$Q^5$ components show the anisotropic peaks, where the corresponding real-space quadrupole configurations are shown in Fig.~\ref{fig: real}(b)--\ref{fig: real}(f), respectively. 
This difference is owing to the fact that the quadrupoles $Q^2$--$Q^5$ belong to the two-dimensional irreducible representation; the sum of the $Q^2$ and $Q^3$ ($Q^4$ and $Q^5$) components leads to the threefold symmetric peak structure. 
It is noted that the appearance of the $\bm{q}=\bm{0}$ component in $S^1_Q(\bm{q})$ is due to the two-dimensional symmetry, where the $Q^1$ belongs to the identity irreducible representation under the $C_{\rm 6v}$ symmetry. 
In the end, this state corresponds to the triple-$Q$ quadrupole state. 
The triple-$Q$ quadrupole state has a similar structure to that in a blue phase in chiral liquid crystals~\cite{Wright_RevModPhys.61.385, Hornreich_PhysRevA.41.1978, fukuda2011quasi, Duzgun_PhysRevE.97.062706} and polar liquid crystals~\cite{Tung_PhysRevB.76.094413, Shamid_PhysRevLett.113.237801}. 
It is noted that the triple-$Q$ structure under the local length constraint in terms of $\bm{u}$ gives rise to the peak intensities at high-harmonic wave vectors in the structure factor, as shown in Figs.~\ref{fig: qspace}(a)--\ref{fig: qspace}(f).

The emergence of the triple-$Q$ quadrupole state is intuitively understood from the competition between the isotropic and chiral biquadratic interactions. 
The isotropic biquadratic interaction tends to favor the quadrupole state with the collinear-type configuration in terms of $\bm{u}_i=(u_i^x, u_i^y, u_i^z)$. 
Meanwhile, the polar biquadratic interaction tends to favor the quadrupole state with the coplanar-type configuration in terms of $\bm{u}_i=(u_i^x, u_i^y, u_i^z)$. 
Such a competition gives rise to the emergence of the single-$Q$ spiral-type configuration in terms of $\bm{u}_i$, which is found in the spin-only model with the isotropic ferromagnetic exchange interaction and the DM interaction.  
In addition, the triple-$Q$ superposition of such a spiral modulation in terms of $\bm{u}_i$ is attributed to the emergence of nonzero contributions from the uniform component in $Q^1$, which mimics the magnetic-field effect in the spin-only model leading to the skyrmion crystal~\cite{rossler2006spontaneous}.

\section{Summary}
\label{sec:summary}

To summarize, we have investigated the symmetry-allowed multipole interactions under crystallographic point groups. 
Based on the representation theory for multipoles, we systematically classify the dipole, quadrupole, octupole, and hexadecapole interactions with both hexagonal and cubic bases. 
The symmetry analysis provides possible multipole interactions for underlying point group symmetries, which can be a source of various multiple-$Q$ multipole orderings. 
To demonstrate that, we construct an effective quadrupole model by starting from the spin-1 model on the noncentrosymmetric triangular lattice, where the DM-type antisymmetric quadrupole interaction is symmetry-allowed. 
By performing the Monte Carlo simulations, we obtain the triple-$Q$ quadrupole state as the ground state by the interplay between the isotropic and antisymmetric quadrupole interactions. 
The present results give a guideline for searching for unknown stabilization mechanisms of multiple-$Q$ multipole orderings. 
In addition, elucidating critical exponents in each multiple-$Q$ model is one of the intriguing research directions. 
Such possibilities of future studies will stimulate future experimental observations of exotic multiple-$Q$ multiple orderings in real materials.

\begin{acknowledgments}
We thank R. Pohle, S. Okumura, Y Kato, and Y. Motome for fruitful discussions.
This research was supported by JSPS KAKENHI Grants Numbers JP21H01037, JP22H04468, JP22H00101, JP22H01183, JP23KJ0557, JP23H04869, JP23K03288, and by JST PRESTO (JPMJPR20L8), JST CREST (JPMJCR23O4), and JST FOREST (JPMJFR2366). 
R.Y. was supported by JSPS Research Fellowship.
Parts of the numerical calculations were performed in the supercomputing systems in ISSP, the University of Tokyo.
\end{acknowledgments}

\appendix

\section{Multipole Interactions under the Cubic Bases for the [100] Bond}
\label{sec: app1}

\subsection{Real Multipole Bases under $m\bar{3}m$}

\begin{table*}
\centering
\caption{\label{tab:cubic_basis}
Cubic bases of the monopole $M^\alpha$, dipoles $D^\alpha$, quadrupoles  $Q^\alpha$, octupoles  $O^\alpha$, and  hexadecapoles  $H^\alpha$.
$\hat{\bm{r}}_i=(x_i,y_i,z_i)$ is the unit vector.
The irreducible representations (Irrep.) of electric and magnetic toroidal multipoles under $O_{\rm h}$ point group, except for the time-reversal property, are also shown; in the case of electric toroidal and magnetic multipoles, the subscript $g/u$ is replaced with $u/g$.
}
\begin{tabular}{ccc}
\hline\hline
rank $l$ & cubic bases & Irrep. \\ \hline
$0$ & $M_i^1=1$ & $A_{1g}$\\
$1$ & $D_i^1=x_i$, $D_i^2=y_i$, $D_i^3=z_i$ & $T_{1u}$\\
$2$ & $Q_i^1=\dfrac{1}{2}(3z_i^2-r_i^2)$, $Q_i^2=\dfrac{\sqrt{3}}{2}(x_i^2-y_i^2)$ & $E_{g}$ \\ 
& $Q_i^3=\sqrt{3}x_iy_i$, $Q_i^4=\sqrt{3}z_ix_i$, $Q_i^5=\sqrt{3}y_iz_i$ & $T_{2g}$\\
$3$ & $O_i^2=\sqrt{15}x_iy_iz_i$ & $A_{2u}$ \\
&   $O_i^4=\dfrac{1}{2}x_i(5x_i^2-3r_i^2)$, $O_i^5=\dfrac{1}{2}y_i(5y_i^2-3r_i^2)$, $O_i^1=\dfrac{1}{2}z_i(5z_i^2-3r_i^2)$ & $T_{1u}$ \\
&$O_i^7=\dfrac{\sqrt{15}}{2}x_i(y_i^2-z_i^2)$,  $O_i^6=\dfrac{\sqrt{15}}{2}y_i(z_i^2-x_i^2)$, $O_i^3=\dfrac{\sqrt{15}}{2}z_i(x_i^2-y_i^2)$ & $T_{2u}$ \\
$4$ & $H_i^1=\dfrac{5\sqrt{21}}{12}\left(x_i^4+y_i^4+z_i^4-\dfrac{3}{5}r_i^4\right)$ & $A_{1g}$ \\
&  $H_i^2=\dfrac{7\sqrt{15}}{6}\left[z_i^4-\dfrac{x_i^4+y_i^4}{2}-\dfrac{3}{7}r_i^2(3z_i^2-r_i^2) \right]$, $H_i^4=\dfrac{7\sqrt{5}}{4}\left[x_i^4-y_i^4-\dfrac{6}{7}r_i^2(x_i^2-y_i^2) \right]$ & $E_{g}$\\
&  $H_i^6=\dfrac{\sqrt{35}}{2}y_iz_i(y_i^2-z_i^2)$, $H_i^7=\dfrac{\sqrt{35}}{2}z_ix_i(z_i^2-x_i^2)$, $H_i^3=\dfrac{\sqrt{35}}{2}x_iy_i(x_i^2-y_i^2)$ & $T_{1g}$\\
 & $H_i^9=\dfrac{\sqrt{5}}{2}y_iz_i(7x_i^2-r_i^2)$, $H_i^8=\dfrac{\sqrt{5}}{2}z_ix_i(7y_i^2-r_i^2)$, $H_i^5=\dfrac{\sqrt{5}}{2}x_iy_i(7z_i^2-r_i^2)$ & $T_{2g}$    
\\ \hline\hline
\end{tabular}
\end{table*}
 
We show the real bases of the multipoles up to $l=4$ for the cubic group $m\bar{3}m$ with the irreducible representation and its subgroups in Table~\ref{tab:cubic_basis}
 
\subsection{Quadrupole Interactions}

\begin{table*}
\centering
\caption{
\label{tab:cubic_Q_AS}
Classification of the antisymmetric quadrupole interactions with the cubic bases for the $[100]$ bond.
The row ``\#" represents the number of the independent interaction parameters.
}
\scalebox{1.0}{
\begin{tabular}{cccccccccccccccc}
\hline \hline 
& $mmm$ & $2mm$ & $m2m$ & $mm2$ & $222$ & $2/m..$ & $.2/m.$ & $..2/m$ & $2..$ & $.2.$ & $..2$ & $m..$ & $.m.$ & $..m$ & $\bar{1}$   \\ \hline
$J^{\mathrm{(AS)}12}_{ij}$ &0 &0 &$\checkmark$ &0 &0 &0 &0 &0 &0 &$\checkmark$ &0 &$\checkmark$ &0 &$\checkmark$ &0 \\
$J^{\mathrm{(AS)}13}_{ij}$ &0 &0 &0 &$\checkmark$ &0 &0 &0 &0 &0 &0 &$\checkmark$ &$\checkmark$ &$\checkmark$ &0 &0 \\
$J^{\mathrm{(AS)}14}_{ij}$ &0 &$\checkmark$ &0 &0 &0 &0 &0 &0 &$\checkmark$ &0 &0 &0 &$\checkmark$ &$\checkmark$ &0 \\
$J^{\mathrm{(AS)}15}_{ij}$ &0 &0 &0 &0 &$\checkmark$ &0 &0 &0 &$\checkmark$ &$\checkmark$ &$\checkmark$ &0 &0 &0 &0 \\
$J^{\mathrm{(AS)}23}_{ij}$ &0 &0 &0 &$\checkmark$ &0 &0 &0 &0 &0 &0 &$\checkmark$ &$\checkmark$ &$\checkmark$ &0 &0 \\
$J^{\mathrm{(AS)}24}_{ij}$ &0 &$\checkmark$ &0 &0 &0 &0 &0 &0 &$\checkmark$ &0 &0 &0 &$\checkmark$ &$\checkmark$ &0 \\
$J^{\mathrm{(AS)}25}_{ij}$ &0 &0 &0 &0 &$\checkmark$ &0 &0 &0 &$\checkmark$ &$\checkmark$ &$\checkmark$ &0 &0 &0 &0 \\
$J^{\mathrm{(AS)}34}_{ij}$ &0 &0 &0 &0 &$\checkmark$ &0 &0 &0 &$\checkmark$ &$\checkmark$ &$\checkmark$ &0 &0 &0 &0 \\
$J^{\mathrm{(AS)}35}_{ij}$ &0 &$\checkmark$ &0 &0 &0 &0 &0 &0 &$\checkmark$ &0 &0 &0 &$\checkmark$ &$\checkmark$ &0 \\
$J^{\mathrm{(AS)}45}_{ij}$ &0 &0 &0 &$\checkmark$ &0 &0 &0 &0 &0 &0 &$\checkmark$ &$\checkmark$ &$\checkmark$ &0 &0 
\\ \hline
\# & 0 &  3 & 1 & 3 & 3 & 0 & 0 & 0 & 6 & 4 & 6 & 4 & 6 & 4 & 0
\\ \hline \hline 
\end{tabular}}
\end{table*}

\begin{table*}
\centering
\caption{
\label{tab:cubic_Q_S}
Classification of the symmetric quadrupole interactions with the cubic bases for the $[100]$ bond.
The row ``\#" represents the number of the independent interaction parameters except for the diagonal components.
}
\scalebox{1.0}{
\begin{tabular}{cccccccccccccccc}
\hline \hline 
& $mmm$ & $2mm$ & $m2m$ & $mm2$ & $222$ & $2/m..$ & $.2/m.$ & $..2/m$ & $2..$ & $.2.$ & $..2$ & $m..$ & $.m.$ & $..m$ & $\bar{1}$   \\ \hline
$J^{\mathrm{(S)}12}_{ij}$ &$\checkmark$ &$\checkmark$ &$\checkmark$ &$\checkmark$ &$\checkmark$ &$\checkmark$ &$\checkmark$ &$\checkmark$ &$\checkmark$ &$\checkmark$ &$\checkmark$ &$\checkmark$ &$\checkmark$ &$\checkmark$ &$\checkmark$ \\
$J^{\mathrm{(S)}13}_{ij}$ &0 &0 &0 &0 &0 &$\checkmark$ &0 &0 &$\checkmark$ &0 &0 &$\checkmark$ &0 &0 &$\checkmark$ \\
$J^{\mathrm{(S)}14}_{ij}$ &0 &0 &0 &0 &0 &0 &0 &$\checkmark$ &0 &0 &$\checkmark$ &0 &0 &$\checkmark$ &$\checkmark$ \\
$J^{\mathrm{(S)}15}_{ij}$ &0 &0 &0 &0 &0 &0 &$\checkmark$ &0 &0 &$\checkmark$ &0 &0 &$\checkmark$ &0 &$\checkmark$ \\
$J^{\mathrm{(S)}23}_{ij}$ &0 &0 &0 &0 &0 &$\checkmark$ &0 &0 &$\checkmark$ &0 &0 &$\checkmark$ &0 &0 &$\checkmark$ \\
$J^{\mathrm{(S)}24}_{ij}$ &0 &0 &0 &0 &0 &0 &0 &$\checkmark$ &0 &0 &$\checkmark$ &0 &0 &$\checkmark$ &$\checkmark$ \\
$J^{\mathrm{(S)}25}_{ij}$ &0 &0 &0 &0 &0 &0 &$\checkmark$ &0 &0 &$\checkmark$ &0 &0 &$\checkmark$ &0 &$\checkmark$ \\
$J^{\mathrm{(S)}34}_{ij}$ &0 &0 &0 &0 &0 &0 &$\checkmark$ &0 &0 &$\checkmark$ &0 &0 &$\checkmark$ &0 &$\checkmark$ \\
$J^{\mathrm{(S)}35}_{ij}$ &0 &0 &0 &0 &0 &0 &0 &$\checkmark$ &0 &0 &$\checkmark$ &0 &0 &$\checkmark$ &$\checkmark$ \\
$J^{\mathrm{(S)}45}_{ij}$ &0 &0 &0 &0 &0 &$\checkmark$ &0 &0 &$\checkmark$ &0 &0 &$\checkmark$ &0 &0 &$\checkmark$ 
\\ \hline
\# & 1 &  1 & 1 & 1 & 1 & 4 & 4 & 4 & 4 & 4 & 4 & 4 & 4 & 4 & 10
\\ \hline \hline 
\end{tabular}}
\end{table*}

We show the classification of the quadrupole interactions with the cubic basis for the [100] bond under the point group symmetries $mmm$, $2mm$, $m2m$, $mm2$, $222$, $2/m..$, $.2/m.$,  $..2/m$, $2..$, $.2.$, $..2$, $m..$, $.m.$, $..m$, and $\bar{1}$. 
The antisymmetric and symmetric quadrupole interactions are presented in Tables~\ref{tab:cubic_Q_AS} and \ref{tab:cubic_Q_S}, respectively. 
In the tables, the 1st, 2nd, and 3rd axes in point groups are $[001]$, $[100]$, and $[010]$, respectively.
We use $\checkmark$ (0) to represent the interaction allowed (forbidden) by the symmetry.

\subsection{Octupole Interactions}

\begin{table*}
\centering
\caption{
\label{tab:cubic_O_AS}
Classification of the antisymmetric octupole interactions with the cubic bases for the $[100]$ bond.
The row ``\#" represents the number of the independent interaction parameters.
}
\scalebox{1.0}{
\begin{tabular}{cccccccccccccccc}
\hline \hline 
& $mmm$ & $2mm$ & $m2m$ & $mm2$ & $222$ & $2/m..$ & $.2/m.$ & $..2/m$ & $2..$ & $.2.$ & $..2$ & $m..$ & $.m.$ & $..m$ & $\bar{1}$   \\ \hline
$J^{\mathrm{(AS)}12}_{ij}$&0	&0	&0	&$\checkmark$&0	&0	&0	&0	&0	&0	&$\checkmark$&$\checkmark$&$\checkmark$&0	&0\\
$J^{\mathrm{(AS)}13}_{ij}$&0	&0	&$\checkmark$&0	&0	&0	&0	&0	&0	&$\checkmark$&0	&$\checkmark$&0	&$\checkmark$&0\\
$J^{\mathrm{(AS)}14}_{ij}$&0	&$\checkmark$&0	&0	&0	&0	&0	&0	&$\checkmark$&0	&0	&0	&$\checkmark$&$\checkmark$&0\\
$J^{\mathrm{(AS)}15}_{ij}$&0	&0	&0	&0	&$\checkmark$&0	&0	&0	&$\checkmark$&$\checkmark$&$\checkmark$&0	&0	&0	&0\\
$J^{\mathrm{(AS)}16}_{ij}$&0	&0	&0	&0	&$\checkmark$&0	&0	&0	&$\checkmark$&$\checkmark$&$\checkmark$&0	&0	&0	&0\\
$J^{\mathrm{(AS)}17}_{ij}$&0	&$\checkmark$&0	&0	&0	&0	&0	&0	&$\checkmark$&0	&0	&0	&$\checkmark$&$\checkmark$&0\\
$J^{\mathrm{(AS)}23}_{ij}$&0	&0	&0	&$\checkmark$&0	&0	&0	&0	&0	&0	&$\checkmark$&$\checkmark$&$\checkmark$&0	&0\\
$J^{\mathrm{(AS)}24}_{ij}$&0	&0	&0	&0	&$\checkmark$&0	&0	&0	&$\checkmark$&$\checkmark$&$\checkmark$&0	&0	&0	&0\\
$J^{\mathrm{(AS)}25}_{ij}$&0	&$\checkmark$&0	&0	&0	&0	&0	&0	&$\checkmark$&0	&0	&0	&$\checkmark$&$\checkmark$&0\\
$J^{\mathrm{(AS)}26}_{ij}$&0	&$\checkmark$&0	&0	&0	&0	&0	&0	&$\checkmark$&0	&0	&0	&$\checkmark$&$\checkmark$&0\\
$J^{\mathrm{(AS)}27}_{ij}$&0	&0	&0	&0	&$\checkmark$&0	&0	&0	&$\checkmark$&$\checkmark$&$\checkmark$&0	&0	&0	&0\\
$J^{\mathrm{(AS)}34}_{ij}$&0	&$\checkmark$&0	&0	&0	&0	&0	&0	&$\checkmark$&0	&0	&0	&$\checkmark$&$\checkmark$&0\\
$J^{\mathrm{(AS)}35}_{ij}$&0	&0	&0	&0	&$\checkmark$&0	&0	&0	&$\checkmark$&$\checkmark$&$\checkmark$&0	&0	&0	&0\\
$J^{\mathrm{(AS)}36}_{ij}$&0	&0	&0	&0	&$\checkmark$&0	&0	&0	&$\checkmark$&$\checkmark$&$\checkmark$&0	&0	&0	&0\\
$J^{\mathrm{(AS)}37}_{ij}$&0	&$\checkmark$&0	&0	&0	&0	&0	&0	&$\checkmark$&0	&0	&0	&$\checkmark$&$\checkmark$&0\\
$J^{\mathrm{(AS)}45}_{ij}$&0	&0	&0	&$\checkmark$&0	&0	&0	&0	&0	&0	&$\checkmark$&$\checkmark$&$\checkmark$&0	&0\\
$J^{\mathrm{(AS)}46}_{ij}$&0	&0	&0	&$\checkmark$&0	&0	&0	&0	&0	&0	&$\checkmark$&$\checkmark$&$\checkmark$&0	&0\\
$J^{\mathrm{(AS)}47}_{ij}$&0	&0	&$\checkmark$&0	&0	&0	&0	&0	&0	&$\checkmark$&0	&$\checkmark$&0	&$\checkmark$&0\\
$J^{\mathrm{(AS)}56}_{ij}$&0	&0	&$\checkmark$&0	&0	&0	&0	&0	&0	&$\checkmark$&0	&$\checkmark$&0	&$\checkmark$&0\\
$J^{\mathrm{(AS)}57}_{ij}$&0	&0	&0	&$\checkmark$&0	&0	&0	&0	&0	&0	&$\checkmark$&$\checkmark$&$\checkmark$&0	&0\\
$J^{\mathrm{(AS)}67}_{ij}$&0	&0	&0	&$\checkmark$&0	&0	&0	&0	&0	&0	&$\checkmark$&$\checkmark$&$\checkmark$&0	&0
\\ \hline
\# & 0 &  6 & 3 & 6 & 6 & 0 & 0 & 0 & 12 & 9 & 12 & 9 & 12 & 9 & 0
\\ \hline \hline 
\end{tabular}
}
\end{table*}

\begin{table*}
\centering
\caption{
\label{tab:cubic_O_S}
Classification of the symmetric octupole interactions with the cubic bases for the $[100]$ bond.
The row ``\#" represents the number of the independent interaction parameters except for the diagonal components.
}
\scalebox{1.0}{
\begin{tabular}{cccccccccccccccc}
\hline \hline 
& $mmm$ & $2mm$ & $m2m$ & $mm2$ & $222$ & $2/m..$ & $.2/m.$ & $..2/m$ & $2..$ & $.2.$ & $..2$ & $m..$ & $.m.$ & $..m$ & $\bar{1}$   \\ \hline
$J^{\mathrm{(S)}12}_{ij}$ &0	&0	&0	&0	&0	&$\checkmark$	&0	&0	&$\checkmark$	&0	&0	&$\checkmark$	&0	&0	&$\checkmark$\\
$J^{\mathrm{(S)}13}_{ij}$&$\checkmark$	&$\checkmark$	&$\checkmark$	&$\checkmark$	&$\checkmark$	&$\checkmark$	&$\checkmark$	&$\checkmark$	&$\checkmark$	&$\checkmark$	&$\checkmark$	&$\checkmark$	&$\checkmark$	&$\checkmark$	&$\checkmark$\\
$J^{\mathrm{(S)}14}_{ij}$&0	&0	&0	&0	&0	&0	&0	&$\checkmark$	&0	&0	&$\checkmark$	&0	&0	&$\checkmark$	&$\checkmark$\\
$J^{\mathrm{(S)}15}_{ij}$&0	&0	&0	&0	&0	&0	&$\checkmark$	&0	&0	&$\checkmark$	&0	&0	&$\checkmark$	&0	&$\checkmark$\\
$J^{\mathrm{(S)}16}_{ij}$&0	&0	&0	&0	&0	&0	&$\checkmark$	&0	&0	&$\checkmark$	&0	&0	&$\checkmark$	&0	&$\checkmark$\\
$J^{\mathrm{(S)}17}_{ij}$&0	&0	&0	&0	&0	&0	&0	&$\checkmark$	&0	&0	&$\checkmark$	&0	&0	&$\checkmark$	&$\checkmark$\\
$J^{\mathrm{(S)}23}_{ij}$&0	&0	&0	&0	&0	&$\checkmark$	&0	&0	&$\checkmark$	&0	&0	&$\checkmark$	&0	&0	&$\checkmark$\\
$J^{\mathrm{(S)}24}_{ij}$&0	&0	&0	&0	&0	&0	&$\checkmark$	&0	&0	&$\checkmark$	&0	&0	&$\checkmark$	&0	&$\checkmark$\\
$J^{\mathrm{(S)}25}_{ij}$&0	&0	&0	&0	&0	&0	&0	&$\checkmark$	&0	&0	&$\checkmark$	&0	&0	&$\checkmark$	&$\checkmark$\\
$J^{\mathrm{(S)}26}_{ij}$&0	&0	&0	&0	&0	&0	&0	&$\checkmark$	&0	&0	&$\checkmark$	&0	&0	&$\checkmark$	&$\checkmark$\\
$J^{\mathrm{(S)}27}_{ij}$&0	&0	&0	&0	&0	&0	&$\checkmark$	&0	&0	&$\checkmark$	&0	&0	&$\checkmark$	&0	&$\checkmark$\\
$J^{\mathrm{(S)}34}_{ij}$&0	&0	&0	&0	&0	&0	&0	&$\checkmark$	&0	&0	&$\checkmark$	&0	&0	&$\checkmark$	&$\checkmark$\\
$J^{\mathrm{(S)}35}_{ij}$&0	&0	&0	&0	&0	&0	&$\checkmark$	&0	&0	&$\checkmark$	&0	&0	&$\checkmark$	&0	&$\checkmark$\\
$J^{\mathrm{(S)}36}_{ij}$&0	&0	&0	&0	&0	&0	&$\checkmark$	&0	&0	&$\checkmark$	&0	&0	&$\checkmark$	&0	&$\checkmark$\\
$J^{\mathrm{(S)}37}_{ij}$&0	&0	&0	&0	&0	&0	&0	&$\checkmark$	&0	&0	&$\checkmark$	&0	&0	&$\checkmark$	&$\checkmark$\\
$J^{\mathrm{(S)}45}_{ij}$&0	&0	&0	&0	&0	&$\checkmark$	&0	&0	&$\checkmark$	&0	&0	&$\checkmark$	&0	&0	&$\checkmark$\\
$J^{\mathrm{(S)}46}_{ij}$&0	&0	&0	&0	&0	&$\checkmark$	&0	&0	&$\checkmark$	&0	&0	&$\checkmark$	&0	&0	&$\checkmark$\\
$J^{\mathrm{(S)}47}_{ij}$&$\checkmark$	&$\checkmark$	&$\checkmark$	&$\checkmark$	&$\checkmark$	&$\checkmark$	&$\checkmark$	&$\checkmark$	&$\checkmark$	&$\checkmark$	&$\checkmark$	&$\checkmark$	&$\checkmark$	&$\checkmark$	&$\checkmark$\\
$J^{\mathrm{(S)}56}_{ij}$&$\checkmark$	&$\checkmark$	&$\checkmark$	&$\checkmark$	&$\checkmark$	&$\checkmark$	&$\checkmark$	&$\checkmark$	&$\checkmark$	&$\checkmark$	&$\checkmark$	&$\checkmark$	&$\checkmark$	&$\checkmark$	&$\checkmark$\\
$J^{\mathrm{(S)}57}_{ij}$&0	&0	&0	&0	&0	&$\checkmark$	&0	&0	&$\checkmark$	&0	&0	&$\checkmark$	&0	&0	&$\checkmark$\\ 
$J^{\mathrm{(S)}67}_{ij}$&0	&0	&0	&0	&0	&$\checkmark$	&0	&0	&$\checkmark$	&0	&0	&$\checkmark$	&0	&0	&$\checkmark$
\\ \hline
\# & 3 &  3 & 3 & 3 & 3 & 9 & 9 & 9 & 9 & 9 & 9 & 9 & 9 & 9 & 21
\\ \hline \hline 
\end{tabular}
}
\end{table*}

We show the classification of the antisymmetric and symmetric octupole interactions with the cubic basis for the [100] bond in Tables~\ref{tab:cubic_O_AS} and \ref{tab:cubic_O_S}, respectively.

\subsection{Hexadecapole Interactions}

\begin{table*}
\centering
\caption{
\label{tab:cubic_H_AS}
Classification of the antisymmetric hexadecapole interactions with the cubic bases for the $[100]$ bond.
The row ``\#" represents the number of the independent interaction parameters.
}
\scalebox{0.9}{
\begin{tabular}{cccccccccccccccc}
\hline \hline 
& $mmm$ & $2mm$ & $m2m$ & $mm2$ & $222$ & $2/m..$ & $.2/m.$ & $..2/m$ & $2..$ & $.2.$ & $..2$ & $m..$ & $.m.$ & $..m$ & $\bar{1}$   \\ \hline
$J^{\mathrm{(AS)}12}_{ij}$&0	&0	&$\checkmark$ &0	&0	&0	&0	&0	&0	&$\checkmark$ &0	&$\checkmark$ &0	&$\checkmark$ &0\\
$J^{\mathrm{(AS)}13}_{ij}$&0	&0	&0	&$\checkmark$ &0	&0	&0	&0	&0	&0	&$\checkmark$ &$\checkmark$ &$\checkmark$ &0	&0\\
$J^{\mathrm{(AS)}14}_{ij}$&0	&0	&$\checkmark$ &0	&0	&0	&0	&0	&0	&$\checkmark$ &0	&$\checkmark$ &0	&$\checkmark$ &0\\
$J^{\mathrm{(AS)}15}_{ij}$&0	&0	&0	&$\checkmark$ &0	&0	&0	&0	&0	&0	&$\checkmark$ &$\checkmark$ &$\checkmark$ &0	&0\\
$J^{\mathrm{(AS)}16}_{ij}$&0	&0	&0	&0	&$\checkmark$ &0	&0	&0	&$\checkmark$ &$\checkmark$ &$\checkmark$ &0	&0	&0	&0\\
$J^{\mathrm{(AS)}17}_{ij}$&0	&$\checkmark$ &0	&0	&0	&0	&0	&0	&$\checkmark$ &0	&0	&0	&$\checkmark$ &$\checkmark$ &0\\
$J^{\mathrm{(AS)}18}_{ij}$&0	&$\checkmark$ &0	&0	&0	&0	&0	&0	&$\checkmark$ &0	&0	&0	&$\checkmark$ &$\checkmark$ &0\\
$J^{\mathrm{(AS)}19}_{ij}$&0	&0	&0	&0	&$\checkmark$ &0	&0	&0	&$\checkmark$ &$\checkmark$ &$\checkmark$ &0	&0	&0	&0\\
$J^{\mathrm{(AS)}23}_{ij}$&0	&0	&0	&$\checkmark$ &0	&0	&0	&0	&0	&0	&$\checkmark$ &$\checkmark$ &$\checkmark$ &0	&0\\
$J^{\mathrm{(AS)}24}_{ij}$&0	&0	&$\checkmark$ &0	&0	&0	&0	&0	&0	&$\checkmark$ &0	&$\checkmark$ &0	&$\checkmark$ &0\\
$J^{\mathrm{(AS)}25}_{ij}$&0	&0	&0	&$\checkmark$ &0	&0	&0	&0	&0	&0	&$\checkmark$ &$\checkmark$ &$\checkmark$ &0	&0\\
$J^{\mathrm{(AS)}26}_{ij}$&0	&0	&0	&0	&$\checkmark$ &0	&0	&0	&$\checkmark$ &$\checkmark$ &$\checkmark$ &0	&0	&0	&0\\
$J^{\mathrm{(AS)}27}_{ij}$&0	&$\checkmark$ &0	&0	&0	&0	&0	&0	&$\checkmark$ &0	&0	&0	&$\checkmark$ &$\checkmark$ &0\\
$J^{\mathrm{(AS)}28}_{ij}$&0	&$\checkmark$ &0	&0	&0	&0	&0	&0	&$\checkmark$ &0	&0	&0	&$\checkmark$ &$\checkmark$ &0\\
$J^{\mathrm{(AS)}29}_{ij}$&0	&0	&0	&0	&$\checkmark$ &0	&0	&0	&$\checkmark$ &$\checkmark$ &$\checkmark$ &0	&0	&0	&0\\
$J^{\mathrm{(AS)}34}_{ij}$&0	&0	&0	&$\checkmark$ &0	&0	&0	&0	&0	&0	&$\checkmark$ &$\checkmark$ &$\checkmark$ &0	&0\\
$J^{\mathrm{(AS)}35}_{ij}$&0	&0	&$\checkmark$ &0	&0	&0	&0	&0	&0	&$\checkmark$ &0	&$\checkmark$ &0	&$\checkmark$ &0\\
$J^{\mathrm{(AS)}36}_{ij}$&0	&$\checkmark$ &0	&0	&0	&0	&0	&0	&$\checkmark$ &0	&0	&0	&$\checkmark$ &$\checkmark$ &0\\
$J^{\mathrm{(AS)}37}_{ij}$&0	&0	&0	&0	&$\checkmark$ &0	&0	&0	&$\checkmark$ &$\checkmark$ &$\checkmark$ &0	&0	&0	&0\\
$J^{\mathrm{(AS)}38}_{ij}$&0	&0	&0	&0	&$\checkmark$ &0	&0	&0	&$\checkmark$ &$\checkmark$ &$\checkmark$ &0	&0	&0	&0\\
$J^{\mathrm{(AS)}39}_{ij}$&0	&$\checkmark$ &0	&0	&0	&0	&0	&0	&$\checkmark$ &0	&0	&0	&$\checkmark$ &$\checkmark$ &0\\
$J^{\mathrm{(AS)}45}_{ij}$&0	&0	&0	&$\checkmark$ &0	&0	&0	&0	&0	&0	&$\checkmark$ &$\checkmark$ &$\checkmark$ &0	&0\\
$J^{\mathrm{(AS)}46}_{ij}$&0	&0	&0	&0	&$\checkmark$ &0	&0	&0	&$\checkmark$ &$\checkmark$ &$\checkmark$ &0	&0	&0	&0\\
$J^{\mathrm{(AS)}47}_{ij}$&0	&$\checkmark$ &0	&0	&0	&0	&0	&0	&$\checkmark$ &0	&0	&0	&$\checkmark$ &$\checkmark$ &0\\
$J^{\mathrm{(AS)}48}_{ij}$&0	&$\checkmark$ &0	&0	&0	&0	&0	&0	&$\checkmark$ &0	&0	&0	&$\checkmark$ &$\checkmark$ &0\\
$J^{\mathrm{(AS)}49}_{ij}$&0	&0	&0	&0	&$\checkmark$ &0	&0	&0	&$\checkmark$ &$\checkmark$ &$\checkmark$ &0	&0	&0	&0\\
$J^{\mathrm{(AS)}56}_{ij}$&0	&$\checkmark$ &0	&0	&0	&0	&0	&0	&$\checkmark$ &0	&0	&0	&$\checkmark$ &$\checkmark$ &0\\
$J^{\mathrm{(AS)}57}_{ij}$&0	&0	&0	&0	&$\checkmark$ &0	&0	&0	&$\checkmark$ &$\checkmark$ &$\checkmark$ &0	&0	&0	&0\\
$J^{\mathrm{(AS)}58}_{ij}$&0	&0	&0	&0	&$\checkmark$ &0	&0	&0	&$\checkmark$ &$\checkmark$ &$\checkmark$ &0	&0	&0	&0\\
$J^{\mathrm{(AS)}59}_{ij}$&0	&$\checkmark$ &0	&0	&0	&0	&0	&0	&$\checkmark$ &0	&0	&0	&$\checkmark$ &$\checkmark$ &0\\
$J^{\mathrm{(AS)}67}_{ij}$&0	&0	&0	&$\checkmark$ &0	&0	&0	&0	&0	&0	&$\checkmark$ &$\checkmark$ &$\checkmark$ &0	&0\\
$J^{\mathrm{(AS)}68}_{ij}$&0	&0	&0	&$\checkmark$ &0	&0	&0	&0	&0	&0	&$\checkmark$ &$\checkmark$ &$\checkmark$ &0	&0\\
$J^{\mathrm{(AS)}69}_{ij}$&0	&0	&$\checkmark$ &0	&0	&0	&0	&0	&0	&$\checkmark$ &0	&$\checkmark$ &0	&$\checkmark$ &0\\
$J^{\mathrm{(AS)}78}_{ij}$&0	&0	&$\checkmark$ &0	&0	&0	&0	&0	&0	&$\checkmark$ &0	&$\checkmark$ &0	&$\checkmark$ &0\\
$J^{\mathrm{(AS)}79}_{ij}$&0	&0	&0	&$\checkmark$ &0	&0	&0	&0	&0	&0	&$\checkmark$ &$\checkmark$ &$\checkmark$ &0	&0\\
$J^{\mathrm{(AS)}89}_{ij}$&0	&0	&0	&$\checkmark$ &0	&0	&0	&0	&0	&0	&$\checkmark$ &$\checkmark$ &$\checkmark$ &0	&0
\\ \hline
\# & 0 &  10 & 6 & 10 & 10 & 0 & 0 & 0 & 20 & 16 & 20 & 16 & 20 & 16 & 0
\\ \hline \hline 
\end{tabular}
}
\end{table*}

\begin{table*}
\centering
\caption{
\label{tab:cubic_H_S}
Classification of the symmetric hexadecapole interactions with the cubic bases for the $[100]$ bond.
The row ``\#" represents the number of the independent interaction parameters except for the diagonal components.
}
\scalebox{0.9}{
\begin{tabular}{cccccccccccccccc}
\hline \hline 
& $mmm$ & $2mm$ & $m2m$ & $mm2$ & $222$ & $2/m..$ & $.2/m.$ & $..2/m$ & $2..$ & $.2.$ & $..2$ & $m..$ & $.m.$ & $..m$ & $\bar{1}$   \\ \hline
$J^{\mathrm{(S)}12}_{ij}$&$\checkmark$ &$\checkmark$ &$\checkmark$ &$\checkmark$ &$\checkmark$ &$\checkmark$ &$\checkmark$ &$\checkmark$ &$\checkmark$ &$\checkmark$ &$\checkmark$ &$\checkmark$ &$\checkmark$ &$\checkmark$ &$\checkmark$ \\
$J^{\mathrm{(S)}13}_{ij}$&0	&0	&0	&0	&0	&$\checkmark$ &0	&0	&$\checkmark$ &0	&0	&$\checkmark$ &0	&0	&$\checkmark$ \\
$J^{\mathrm{(S)}14}_{ij}$&$\checkmark$ &$\checkmark$ &$\checkmark$ &$\checkmark$ &$\checkmark$ &$\checkmark$ &$\checkmark$ &$\checkmark$ &$\checkmark$ &$\checkmark$ &$\checkmark$ &$\checkmark$ &$\checkmark$ &$\checkmark$ &$\checkmark$ \\
$J^{\mathrm{(S)}15}_{ij}$&0	&0	&0	&0	&0	&$\checkmark$ &0	&0	&$\checkmark$ &0	&0	&$\checkmark$ &0	&0	&$\checkmark$ \\
$J^{\mathrm{(S)}16}_{ij}$&0	&0	&0	&0	&0	&0	&$\checkmark$ &0	&0	&$\checkmark$ &0	&0	&$\checkmark$ &0	&$\checkmark$ \\
$J^{\mathrm{(S)}17}_{ij}$&0	&0	&0	&0	&0	&0	&0	&$\checkmark$ &0	&0	&$\checkmark$ &0	&0	&$\checkmark$ &$\checkmark$ \\
$J^{\mathrm{(S)}18}_{ij}$&0	&0	&0	&0	&0	&0	&0	&$\checkmark$ &0	&0	&$\checkmark$ &0	&0	&$\checkmark$ &$\checkmark$ \\
$J^{\mathrm{(S)}19}_{ij}$&0	&0	&0	&0	&0	&0	&$\checkmark$ &0	&0	&$\checkmark$ &0	&0	&$\checkmark$ &0	&$\checkmark$ \\
$J^{\mathrm{(S)}23}_{ij}$&0	&0	&0	&0	&0	&$\checkmark$ &0	&0	&$\checkmark$ &0	&0	&$\checkmark$ &0	&0	&$\checkmark$ \\
$J^{\mathrm{(S)}24}_{ij}$&$\checkmark$ &$\checkmark$ &$\checkmark$ &$\checkmark$ &$\checkmark$ &$\checkmark$ &$\checkmark$ &$\checkmark$ &$\checkmark$ &$\checkmark$ &$\checkmark$ &$\checkmark$ &$\checkmark$ &$\checkmark$ &$\checkmark$ \\
$J^{\mathrm{(S)}25}_{ij}$&0	&0	&0	&0	&0	&$\checkmark$ &0	&0	&$\checkmark$ &0	&0	&$\checkmark$ &0	&0	&$\checkmark$ \\
$J^{\mathrm{(S)}26}_{ij}$&0	&0	&0	&0	&0	&0	&$\checkmark$ &0	&0	&$\checkmark$ &0	&0	&$\checkmark$ &0	&$\checkmark$ \\
$J^{\mathrm{(S)}27}_{ij}$&0	&0	&0	&0	&0	&0	&0	&$\checkmark$ &0	&0	&$\checkmark$ &0	&0	&$\checkmark$ &$\checkmark$ \\
$J^{\mathrm{(S)}28}_{ij}$&0	&0	&0	&0	&0	&0	&0	&$\checkmark$ &0	&0	&$\checkmark$ &0	&0	&$\checkmark$ &$\checkmark$ \\
$J^{\mathrm{(S)}29}_{ij}$&0	&0	&0	&0	&0	&0	&$\checkmark$ &0	&0	&$\checkmark$ &0	&0	&$\checkmark$ &0	&$\checkmark$ \\
$J^{\mathrm{(S)}34}_{ij}$&0	&0	&0	&0	&0	&$\checkmark$ &0	&0	&$\checkmark$ &0	&0	&$\checkmark$ &0	&0	&$\checkmark$ \\
$J^{\mathrm{(S)}35}_{ij}$&$\checkmark$ &$\checkmark$ &$\checkmark$ &$\checkmark$ &$\checkmark$ &$\checkmark$ &$\checkmark$ &$\checkmark$ &$\checkmark$ &$\checkmark$ &$\checkmark$ &$\checkmark$ &$\checkmark$ &$\checkmark$ &$\checkmark$ \\
$J^{\mathrm{(S)}36}_{ij}$&0	&0	&0	&0	&0	&0	&0	&$\checkmark$ &0	&0	&$\checkmark$ &0	&0	&$\checkmark$ &$\checkmark$ \\
$J^{\mathrm{(S)}37}_{ij}$&0	&0	&0	&0	&0	&0	&$\checkmark$ &0	&0	&$\checkmark$ &0	&0	&$\checkmark$ &0	&$\checkmark$ \\
$J^{\mathrm{(S)}38}_{ij}$&0	&0	&0	&0	&0	&0	&$\checkmark$ &0	&0	&$\checkmark$ &0	&0	&$\checkmark$ &0	&$\checkmark$ \\
$J^{\mathrm{(S)}39}_{ij}$&0	&0	&0	&0	&0	&0	&0	&$\checkmark$ &0	&0	&$\checkmark$ &0	&0	&$\checkmark$ &$\checkmark$ \\
$J^{\mathrm{(S)}45}_{ij}$&0	&0	&0	&0	&0	&$\checkmark$ &0	&0	&$\checkmark$ &0	&0	&$\checkmark$ &0	&0	&$\checkmark$ \\
$J^{\mathrm{(S)}46}_{ij}$&0	&0	&0	&0	&0	&0	&$\checkmark$ &0	&0	&$\checkmark$ &0	&0	&$\checkmark$ &0	&$\checkmark$ \\
$J^{\mathrm{(S)}47}_{ij}$&0	&0	&0	&0	&0	&0	&0	&$\checkmark$ &0	&0	&$\checkmark$ &0	&0	&$\checkmark$ &$\checkmark$ \\
$J^{\mathrm{(S)}48}_{ij}$&0	&0	&0	&0	&0	&0	&0	&$\checkmark$ &0	&0	&$\checkmark$ &0	&0	&$\checkmark$ &$\checkmark$ \\
$J^{\mathrm{(S)}49}_{ij}$&0	&0	&0	&0	&0	&0	&$\checkmark$ &0	&0	&$\checkmark$ &0	&0	&$\checkmark$ &0	&$\checkmark$ \\
$J^{\mathrm{(S)}56}_{ij}$&0	&0	&0	&0	&0	&0	&0	&$\checkmark$ &0	&0	&$\checkmark$ &0	&0	&$\checkmark$ &$\checkmark$ \\
$J^{\mathrm{(S)}57}_{ij}$&0	&0	&0	&0	&0	&0	&$\checkmark$ &0	&0	&$\checkmark$ &0	&0	&$\checkmark$ &0	&$\checkmark$ \\
$J^{\mathrm{(S)}58}_{ij}$&0	&0	&0	&0	&0	&0	&$\checkmark$ &0	&0	&$\checkmark$ &0	&0	&$\checkmark$ &0	&$\checkmark$ \\
$J^{\mathrm{(S)}59}_{ij}$&0	&0	&0	&0	&0	&0	&0	&$\checkmark$ &0	&0	&$\checkmark$ &0	&0	&$\checkmark$ &$\checkmark$ \\
$J^{\mathrm{(S)}67}_{ij}$&0	&0	&0	&0	&0	&$\checkmark$ &0	&0	&$\checkmark$ &0	&0	&$\checkmark$ &0	&0	&$\checkmark$ \\
$J^{\mathrm{(S)}68}_{ij}$&0	&0	&0	&0	&0	&$\checkmark$ &0	&0	&$\checkmark$ &0	&0	&$\checkmark$ &0	&0	&$\checkmark$ \\
$J^{\mathrm{(S)}69}_{ij}$&$\checkmark$ &$\checkmark$ &$\checkmark$ &$\checkmark$ &$\checkmark$ &$\checkmark$ &$\checkmark$ &$\checkmark$ &$\checkmark$ &$\checkmark$ &$\checkmark$ &$\checkmark$ &$\checkmark$ &$\checkmark$ &$\checkmark$ \\
$J^{\mathrm{(S)}78}_{ij}$&$\checkmark$ &$\checkmark$ &$\checkmark$ &$\checkmark$ &$\checkmark$ &$\checkmark$ &$\checkmark$ &$\checkmark$ &$\checkmark$ &$\checkmark$ &$\checkmark$ &$\checkmark$ &$\checkmark$ &$\checkmark$ &$\checkmark$ \\
$J^{\mathrm{(S)}79}_{ij}$&0	&0	&0	&0	&0	&$\checkmark$ &0	&0	&$\checkmark$ &0	&0	&$\checkmark$ &0	&0	&$\checkmark$ \\
$J^{\mathrm{(S)}89}_{ij}$&0	&0	&0	&0	&0	&$\checkmark$ &0	&0	&$\checkmark$ &0	&0	&$\checkmark$ &0	&0	&$\checkmark$
\\ \hline
\# & 6 &  6 & 6 & 6 & 6 & 16 & 16 & 16 & 16 & 16 & 16 & 16 & 16 & 16 & 36
\\ \hline \hline 
\end{tabular}
}
\end{table*}

We show the classification of the antisymmetric and symmetric hexadecapole interactions with the cubic basis for the [100] bond in Tables~\ref{tab:cubic_H_AS} and \ref{tab:cubic_H_S}, respectively.

\section{Multipole Interactions under the Hexagonal Bases for the [001] Bond}
\label{sec: app2}

The multipole interactions for the [001] bond for the hexagonal bases in Table~\ref{tab:hexagonal_basis} are classified under the point group symmetries $6/mmm$, $\bar{6}m2$, $\bar{6}2m$, $6mm$, $622$, $6/m..$, $6..$, $\bar{6}..$, $\bar{3}m1$, $\bar{3}1m$, $3m1$, $31m$, $321$, $312$, $3..$, $\bar{3}..$, and $mmm$.

\subsection{Quadrupole Interaction}

\begin{table*}
\centering
\caption{
\label{tab:hexagonal_Q_AS2}
Classification of the antisymmetric quadrupole interactions with the hexagonal bases for the $[001]$ bond.
The row ``\#" represents the number of the independent interaction parameters.
}
\scalebox{0.85}{
\begin{tabular}{ccccccccccccccccccc}
\hline \hline
& $6/mmm$ & $\bar{6}m2$ & $\bar{6}2m$ & $6mm$ & $622$ & $6/m..$ & $6..$ & $\bar{6}..$ & $\bar{3}m1$ & $\bar{3}1m$ & $3m1$ & $31m$ & $321$ & $312$ & $3..$  & $\bar{3}..$ & $mmm$   \\ \hline
$J^{\mathrm{(AS)}12}_{ij}$&0	&0	&0	&0	&0	&0	&0	&0	&0	&0	&0	&0	&0	&0	&0	&0	&0\\
$J^{\mathrm{(AS)}13}_{ij}$&0	&0	&0	&0	&0	&0	&0	&0	&0	&0	&0	&0	&0	&0	&0	&0	&0\\
$J^{\mathrm{(AS)}14}_{ij}$&0	&0	&0	&0	&0	&0	&0	&0	&0	&0	&0	&0	&0	&0	&0	&0	&0\\
$J^{\mathrm{(AS)}15}_{ij}$&0	&0	&0	&0	&0	&0	&0	&0	&0	&0	&0	&0	&0	&0	&0	&0	&0\\
$J^{\mathrm{(AS)}23}_{ij}$&0	&0	&0	&0	&$\checkmark$&0	&$\checkmark$&0	&0	&0	&0	&0	&$\checkmark$&$\checkmark$&$\checkmark$&0	&0\\
$J^{\mathrm{(AS)}24}_{ij}$&0	&$\checkmark$&0	&0	&0	&0	&0	&$\checkmark$&0	&0	&$\checkmark$&0	&0	&$\checkmark$&$\checkmark$&0	&0\\
$J^{\mathrm{(AS)}25}_{ij}$&0	&0	&$\checkmark$&0	&0	&0	&0	&$\checkmark$&0	&0	&0	&$\checkmark$&$\checkmark$&0	&$\checkmark$&0	&0\\
$J^{\mathrm{(AS)}34}_{ij}$&0	&0	&$\checkmark$&0	&0	&0	&0	&$\checkmark$&0	&0	&0	&$\checkmark$&$\checkmark$&0	&$\checkmark$&0	&0\\
$J^{\mathrm{(AS)}35}_{ij}$&0	&$\checkmark$&0	&0	&0	&0	&0	&$\checkmark$&0	&0	&$\checkmark$&0	&0	&$\checkmark$&$\checkmark$&0	&0\\
$J^{\mathrm{(AS)}45}_{ij}$&0	&0	&0	&0	&$\checkmark$&0	&$\checkmark$&0	&0	&0	&0	&0	&$\checkmark$&$\checkmark$&$\checkmark$&0	&0
\\ \hline
\# & 0 &  2 & 2 & 0 & 2 & 0 & 2 & 4 & 0 & 0 & 2 & 2 & 4 & 4 & 6 & 0 & 0
\\ \hline \hline 
\end{tabular}
}
\end{table*}

\begin{table*}
\centering
\caption{
\label{tab:hexagonal_Q_S2}
Classification of the symmetric quadrupole interactions with the hexagonal bases for the $[001]$ bond.
The row ``\#" represents the number of the independent interaction parameters except for the diagonal components.
}
\scalebox{0.85}{
\begin{tabular}{ccccccccccccccccccc}
\hline \hline
& $6/mmm$ & $\bar{6}m2$ & $\bar{6}2m$ & $6mm$ & $622$ & $6/m..$ & $6..$ & $\bar{6}..$ & $\bar{3}m1$ & $\bar{3}1m$ & $3m1$ & $31m$ & $321$ & $312$ & $3..$  & $\bar{3}..$ & $mmm$   \\ \hline
$J^{\mathrm{(S)}12}_{ij}$&0	&0	&0	&0	&0	&0	&0	&0	&0	&0	&0	&0	&0	&0	&0	&0	&0\\
$J^{\mathrm{(S)}13}_{ij}$&0	&0	&0	&0	&0	&0	&0	&0	&0	&0	&0	&0	&0	&0	&0	&0	&0\\
$J^{\mathrm{(S)}14}_{ij}$&0	&0	&0	&0	&0	&0	&0	&0	&0	&0	&0	&0	&0	&0	&0	&0	&0\\
$J^{\mathrm{(S)}15}_{ij}$&0	&0	&0	&0	&0	&0	&0	&0	&0	&0	&0	&0	&0	&0	&0	&0	&$\checkmark$\\
$J^{\mathrm{(S)}23}_{ij}$&0	&0	&0	&0	&0	&0	&0	&0	&0	&0	&0	&0	&0	&0	&0	&0	&0\\
$J^{\mathrm{(S)}24}_{ij}$&0	&0	&0	&0	&0	&0	&0	&0	&$\checkmark$&0	&$\checkmark$&0	&$\checkmark$&0	&$\checkmark$&$\checkmark$&0\\
$J^{\mathrm{(S)}25}_{ij}$&0	&0	&0	&0	&0	&0	&0	&0	&0	&$\checkmark$&0	&$\checkmark$&0	&$\checkmark$&$\checkmark$&$\checkmark$&0\\
$J^{\mathrm{(S)}34}_{ij}$&0	&0	&0	&0	&0	&0	&0	&0	&0	&$\checkmark$&0	&$\checkmark$&0	&$\checkmark$&$\checkmark$&$\checkmark$&0\\
$J^{\mathrm{(S)}35}_{ij}$&0	&0	&0	&0	&0	&0	&0	&0	&$\checkmark$&0	&$\checkmark$&0	&$\checkmark$&0	&$\checkmark$&$\checkmark$&0\\
$J^{\mathrm{(S)}45}_{ij}$&0	&0	&0	&0	&0	&0	&0	&0	&0	&0	&0	&0	&0	&0	&0	&0	&0
\\ \hline
\# & 0 &  0 & 0 & 0 & 0 & 0 & 0 & 0 & 2 & 2 & 2 & 2 & 2 & 2 & 4 & 4 & 1
\\ \hline \hline 
\end{tabular}
}
\end{table*}

We show the classification of the antisymmetric and symmetric quadrupole interactions with the hexagonal basis for the [001] bond in Tables~\ref{tab:hexagonal_Q_AS2} and \ref{tab:hexagonal_Q_S2}, respectively.

\subsection{Octupole Interaction}

\begin{table*}
\centering
\caption{
\label{tab:hexagonal_O_AS2}
Classification of the antisymmetric octupole interactions with the hexagonal bases for the $[001]$ bond.
The row ``\#" represents the number of the independent interaction parameters.
}
\scalebox{0.85}{
\begin{tabular}{ccccccccccccccccccc}
\hline \hline
& $6/mmm$ & $\bar{6}m2$ & $\bar{6}2m$ & $6mm$ & $622$ & $6/m..$ & $6..$ & $\bar{6}..$ & $\bar{3}m1$ & $\bar{3}1m$ & $3m1$ & $31m$ & $321$ & $312$ & $3..$  & $\bar{3}..$ & $mmm$   \\ \hline
$J^{\mathrm{(AS)}12}_{ij}$&0	&0	&$\checkmark$&0	&0	&0	&0	&$\checkmark$&0	&0	&0	&$\checkmark$&$\checkmark$&0	&$\checkmark$&0	&0\\
$J^{\mathrm{(AS)}13}_{ij}$&0	&$\checkmark$&0	&0	&0	&0	&0	&$\checkmark$&0	&0	&$\checkmark$&0	&0	&$\checkmark$&$\checkmark$&0	&0\\
$J^{\mathrm{(AS)}14}_{ij}$&0	&0	&0	&0	&0	&0	&0	&0	&0	&0	&0	&0	&0	&0	&0	&0	&0\\
$J^{\mathrm{(AS)}15}_{ij}$&0	&0	&0	&0	&0	&0	&0	&0	&0	&0	&0	&0	&0	&0	&0	&0	&0\\
$J^{\mathrm{(AS)}16}_{ij}$&0	&0	&0	&0	&0	&0	&0	&0	&0	&0	&0	&0	&0	&0	&0	&0	&0\\
$J^{\mathrm{(AS)}17}_{ij}$&0	&0	&0	&0	&0	&0	&0	&0	&0	&0	&0	&0	&0	&0	&0	&0	&0\\
$J^{\mathrm{(AS)}23}_{ij}$&0	&0	&0	&0	&$\checkmark$&0	&$\checkmark$&0	&0	&0	&0	&0	&$\checkmark$&$\checkmark$&$\checkmark$&0	&0\\
$J^{\mathrm{(AS)}24}_{ij}$&0	&0	&0	&0	&0	&0	&0	&0	&0	&0	&0	&0	&0	&0	&0	&0	&0\\
$J^{\mathrm{(AS)}25}_{ij}$&0	&0	&0	&0	&0	&0	&0	&0	&0	&0	&0	&0	&0	&0	&0	&0	&0\\
$J^{\mathrm{(AS)}26}_{ij}$&0	&0	&0	&0	&0	&0	&0	&0	&0	&0	&0	&0	&0	&0	&0	&0	&0\\
$J^{\mathrm{(AS)}27}_{ij}$&0	&0	&0	&0	&0	&0	&0	&0	&0	&0	&0	&0	&0	&0	&0	&0	&0\\
$J^{\mathrm{(AS)}34}_{ij}$&0	&0	&0	&0	&0	&0	&0	&0	&0	&0	&0	&0	&0	&0	&0	&0	&0\\
$J^{\mathrm{(AS)}35}_{ij}$&0	&0	&0	&0	&0	&0	&0	&0	&0	&0	&0	&0	&0	&0	&0	&0	&0\\
$J^{\mathrm{(AS)}36}_{ij}$&0	&0	&0	&0	&0	&0	&0	&0	&0	&0	&0	&0	&0	&0	&0	&0	&0\\
$J^{\mathrm{(AS)}37}_{ij}$&0	&0	&0	&0	&0	&0	&0	&0	&0	&0	&0	&0	&0	&0	&0	&0	&0\\
$J^{\mathrm{(AS)}45}_{ij}$&0	&0	&0	&0	&$\checkmark$&0	&$\checkmark$&0	&0	&0	&0	&0	&$\checkmark$&$\checkmark$&$\checkmark$&0	&0\\
$J^{\mathrm{(AS)}46}_{ij}$&0	&$\checkmark$&0	&0	&0	&0	&0	&$\checkmark$&0	&0	&$\checkmark$&0	&0	&$\checkmark$&$\checkmark$&0	&0\\
$J^{\mathrm{(AS)}47}_{ij}$&0	&0	&$\checkmark$&0	&0	&0	&0	&$\checkmark$&0	&0	&0	&$\checkmark$&$\checkmark$&0	&$\checkmark$&0	&0\\
$J^{\mathrm{(AS)}56}_{ij}$&0	&0	&$\checkmark$&0	&0	&0	&0	&$\checkmark$&0	&0	&0	&$\checkmark$&$\checkmark$&0	&$\checkmark$&0	&0\\
$J^{\mathrm{(AS)}57}_{ij}$&0	&$\checkmark$&0	&0	&0	&0	&0	&$\checkmark$&0	&0	&$\checkmark$&0	&0	&$\checkmark$&$\checkmark$&0	&0\\
$J^{\mathrm{(AS)}67}_{ij}$&0	&0	&0	&0	&$\checkmark$&0	&$\checkmark$&0	&0	&0	&0	&0	&$\checkmark$&$\checkmark$&$\checkmark$&0	&0
\\ \hline
\# & 0 &  3 & 3 & 0 & 3 & 0 & 3 & 6 & 0 & 0 & 3 & 3 & 6 & 6 & 9 & 0 & 0
\\ \hline \hline 
\end{tabular}
}
\end{table*}

\begin{table*}
\centering
\caption{
\label{tab:hexagonal_O_S2}
Classification of the symmetric octupole interactions with the hexagonal bases for the $[001]$ bond.
The row ``\#" represents the number of the independent interaction parameters except for the diagonal components.
}
\scalebox{0.85}{
\begin{tabular}{ccccccccccccccccccc}
\hline \hline
& $6/mmm$ & $\bar{6}m2$ & $\bar{6}2m$ & $6mm$ & $622$ & $6/m..$ & $6..$ & $\bar{6}..$ & $\bar{3}m1$ & $\bar{3}1m$ & $3m1$ & $31m$ & $321$ & $312$ & $3..$  & $\bar{3}..$ & $mmm$   \\ \hline
$J^{\mathrm{(S)}12}_{ij}$&0	&0	&0	&0	&0	&0	&0	&0	&0	&$\checkmark$&0	&$\checkmark$&0	&$\checkmark$&$\checkmark$&$\checkmark$&0\\
$J^{\mathrm{(S)}13}_{ij}$&0	&0	&0	&0	&0	&0	&0	&0	&$\checkmark$&0	&$\checkmark$&0	&$\checkmark$&0	&$\checkmark$&$\checkmark$&0\\
$J^{\mathrm{(S)}14}_{ij}$&0	&0	&0	&0	&0	&0	&0	&0	&0	&0	&0	&0	&0	&0	&0	&0	&0\\
$J^{\mathrm{(S)}15}_{ij}$&0	&0	&0	&0	&0	&0	&0	&0	&0	&0	&0	&0	&0	&0	&0	&0	&0\\
$J^{\mathrm{(S)}16}_{ij}$&0	&0	&0	&0	&0	&0	&0	&0	&0	&0	&0	&0	&0	&0	&0	&0	&0\\
$J^{\mathrm{(S)}17}_{ij}$&0	&0	&0	&0	&0	&0	&0	&0	&0	&0	&0	&0	&0	&0	&0	&0	&$\checkmark$\\
$J^{\mathrm{(S)}23}_{ij}$&0	&0	&0	&0	&0	&$\checkmark$&$\checkmark$&$\checkmark$&0	&0	&0	&0	&0	&0	&$\checkmark$&$\checkmark$&0\\
$J^{\mathrm{(S)}24}_{ij}$&0	&0	&0	&0	&0	&0	&0	&0	&0	&0	&0	&0	&0	&0	&0	&0	&$\checkmark$\\
$J^{\mathrm{(S)}25}_{ij}$&0	&0	&0	&0	&0	&0	&0	&0	&0	&0	&0	&0	&0	&0	&0	&0	&0\\
$J^{\mathrm{(S)}26}_{ij}$&0	&0	&0	&0	&0	&0	&0	&0	&0	&0	&0	&0	&0	&0	&0	&0	&0\\
$J^{\mathrm{(S)}27}_{ij}$&0	&0	&0	&0	&0	&0	&0	&0	&0	&0	&0	&0	&0	&0	&0	&0	&0\\
$J^{\mathrm{(S)}34}_{ij}$&0	&0	&0	&0	&0	&0	&0	&0	&0	&0	&0	&0	&0	&0	&0	&0	&0\\
$J^{\mathrm{(S)}35}_{ij}$&0	&0	&0	&0	&0	&0	&0	&0	&0	&0	&0	&0	&0	&0	&0	&0	&$\checkmark$\\
$J^{\mathrm{(S)}36}_{ij}$&0	&0	&0	&0	&0	&0	&0	&0	&0	&0	&0	&0	&0	&0	&0	&0	&0\\
$J^{\mathrm{(S)}37}_{ij}$&0	&0	&0	&0	&0	&0	&0	&0	&0	&0	&0	&0	&0	&0	&0	&0	&0\\
$J^{\mathrm{(S)}45}_{ij}$&0	&0	&0	&0	&0	&0	&0	&0	&0	&0	&0	&0	&0	&0	&0	&0	&0\\
$J^{\mathrm{(S)}46}_{ij}$&0	&0	&0	&0	&0	&0	&0	&0	&$\checkmark$&0	&$\checkmark$&0	&$\checkmark$&0	&$\checkmark$&$\checkmark$&0\\
$J^{\mathrm{(S)}47}_{ij}$&0	&0	&0	&0	&0	&0	&0	&0	&0	&$\checkmark$&0	&$\checkmark$&0	&$\checkmark$&$\checkmark$&$\checkmark$&0\\
$J^{\mathrm{(S)}56}_{ij}$&0	&0	&0	&0	&0	&0	&0	&0	&0	&$\checkmark$&0	&$\checkmark$&0	&$\checkmark$&$\checkmark$&$\checkmark$&0\\
$J^{\mathrm{(S)}57}_{ij}$&0	&0	&0	&0	&0	&0	&0	&0	&$\checkmark$&0	&$\checkmark$&0	&$\checkmark$&0	&$\checkmark$&$\checkmark$&0\\
$J^{\mathrm{(S)}67}_{ij}$&0	&0	&0	&0	&0	&0	&0	&0	&0	&0	&0	&0	&0	&0	&0	&0	&0
\\ \hline
\# & 0 &  0 & 0 & 0 & 0 & 1 & 1 & 1 & 3 & 3 & 3 & 3 & 3 & 3 & 7 & 7 & 3
\\ \hline \hline 
\end{tabular}
}
\end{table*}

We show the classification of the antisymmetric and symmetric octupole interactions with the hexagonal basis for the [001] bond in Tables~\ref{tab:hexagonal_O_AS2} and \ref{tab:hexagonal_O_S2}, respectively.

\subsection{Hexadecapole Interaction}

\begin{table*}
\centering
\caption{
\label{tab:hexagonal_H_AS2}
Classification of the antisymmetric hexadecapole interactions with the hexagonal bases for the $[001]$ bond.
The row ``\#" represents the number of the independent interaction parameters.
}
\scalebox{0.85}{
\begin{tabular}{ccccccccccccccccccc}
\hline \hline
& $6/mmm$ & $\bar{6}m2$ & $\bar{6}2m$ & $6mm$ & $622$ & $6/m..$ & $6..$ & $\bar{6}..$ & $\bar{3}m1$ & $\bar{3}1m$ & $3m1$ & $31m$ & $321$ & $312$ & $3..$  & $\bar{3}..$ & $mmm$   \\ \hline
$J^{\mathrm{(AS)}12}_{ij}$&0	&$\checkmark$&0	&0	&0	&0	&0	&$\checkmark$&0	&0	&$\checkmark$&0	&0	&$\checkmark$&$\checkmark$&0	&0\\
$J^{\mathrm{(AS)}13}_{ij}$&0	&0	&$\checkmark$&0	&0	&0	&0	&$\checkmark$&0	&0	&0	&$\checkmark$&$\checkmark$&0	&$\checkmark$&0	&0\\
$J^{\mathrm{(AS)}14}_{ij}$&0	&0	&0	&0	&0	&0	&0	&0	&0	&0	&0	&0	&0	&0	&0	&0	&0\\
$J^{\mathrm{(AS)}15}_{ij}$&0	&0	&0	&0	&0	&0	&0	&0	&0	&0	&0	&0	&0	&0	&0	&0	&0\\
$J^{\mathrm{(AS)}16}_{ij}$&0	&0	&0	&0	&0	&0	&0	&0	&0	&0	&0	&0	&0	&0	&0	&0	&0\\
$J^{\mathrm{(AS)}17}_{ij}$&0	&0	&0	&0	&0	&0	&0	&0	&0	&0	&0	&0	&0	&0	&0	&0	&0\\
$J^{\mathrm{(AS)}18}_{ij}$&0	&0	&0	&0	&0	&0	&0	&0	&0	&0	&0	&0	&0	&0	&0	&0	&0\\
$J^{\mathrm{(AS)}19}_{ij}$&0	&0	&0	&0	&0	&0	&0	&0	&0	&0	&0	&0	&0	&0	&0	&0	&0\\
$J^{\mathrm{(AS)}23}_{ij}$&0	&0	&0	&0	&$\checkmark$&0	&$\checkmark$&0	&0	&0	&0	&0	&$\checkmark$&$\checkmark$&$\checkmark$&0	&0\\
$J^{\mathrm{(AS)}24}_{ij}$&0	&0	&0	&0	&0	&0	&0	&0	&0	&0	&0	&0	&0	&0	&0	&0	&0\\
$J^{\mathrm{(AS)}25}_{ij}$&0	&0	&0	&0	&0	&0	&0	&0	&0	&0	&0	&0	&0	&0	&0	&0	&0\\
$J^{\mathrm{(AS)}26}_{ij}$&0	&0	&0	&0	&0	&0	&0	&0	&0	&0	&0	&0	&0	&0	&0	&0	&0\\
$J^{\mathrm{(AS)}27}_{ij}$&0	&0	&0	&0	&0	&0	&0	&0	&0	&0	&0	&0	&0	&0	&0	&0	&0\\
$J^{\mathrm{(AS)}28}_{ij}$&0	&0	&0	&0	&0	&0	&0	&0	&0	&0	&0	&0	&0	&0	&0	&0	&0\\
$J^{\mathrm{(AS)}29}_{ij}$&0	&0	&0	&0	&0	&0	&0	&0	&0	&0	&0	&0	&0	&0	&0	&0	&0\\
$J^{\mathrm{(AS)}34}_{ij}$&0	&0	&0	&0	&0	&0	&0	&0	&0	&0	&0	&0	&0	&0	&0	&0	&0\\
$J^{\mathrm{(AS)}35}_{ij}$&0	&0	&0	&0	&0	&0	&0	&0	&0	&0	&0	&0	&0	&0	&0	&0	&0\\
$J^{\mathrm{(AS)}36}_{ij}$&0	&0	&0	&0	&0	&0	&0	&0	&0	&0	&0	&0	&0	&0	&0	&0	&0\\
$J^{\mathrm{(AS)}37}_{ij}$&0	&0	&0	&0	&0	&0	&0	&0	&0	&0	&0	&0	&0	&0	&0	&0	&0\\
$J^{\mathrm{(AS)}38}_{ij}$&0	&0	&0	&0	&0	&0	&0	&0	&0	&0	&0	&0	&0	&0	&0	&0	&0\\
$J^{\mathrm{(AS)}39}_{ij}$&0	&0	&0	&0	&0	&0	&0	&0	&0	&0	&0	&0	&0	&0	&0	&0	&0\\
$J^{\mathrm{(AS)}45}_{ij}$&0	&0	&0	&0	&$\checkmark$&0	&$\checkmark$&0	&0	&0	&0	&0	&$\checkmark$&$\checkmark$&$\checkmark$&0	&0\\
$J^{\mathrm{(AS)}46}_{ij}$&0	&0	&$\checkmark$&0	&0	&0	&0	&$\checkmark$&0	&0	&0	&$\checkmark$&$\checkmark$&0	&$\checkmark$&0	&0\\
$J^{\mathrm{(AS)}47}_{ij}$&0	&$\checkmark$&0	&0	&0	&0	&0	&$\checkmark$&0	&0	&$\checkmark$&0	&0	&$\checkmark$&$\checkmark$&0	&0\\
$J^{\mathrm{(AS)}48}_{ij}$&0	&$\checkmark$&0	&0	&0	&0	&0	&$\checkmark$&0	&0	&$\checkmark$&0	&0	&$\checkmark$&$\checkmark$&0	&0\\
$J^{\mathrm{(AS)}49}_{ij}$&0	&0	&$\checkmark$&0	&0	&0	&0	&$\checkmark$&0	&0	&0	&$\checkmark$&$\checkmark$&0	&$\checkmark$&0	&0\\
$J^{\mathrm{(AS)}56}_{ij}$&0	&$\checkmark$&0	&0	&0	&0	&0	&$\checkmark$&0	&0	&$\checkmark$&0	&0	&$\checkmark$&$\checkmark$&0	&0\\
$J^{\mathrm{(AS)}57}_{ij}$&0	&0	&$\checkmark$&0	&0	&0	&0	&$\checkmark$&0	&0	&0	&$\checkmark$&$\checkmark$&0	&$\checkmark$&0	&0\\
$J^{\mathrm{(AS)}58}_{ij}$&0	&0	&$\checkmark$&0	&0	&0	&0	&$\checkmark$&0	&0	&0	&$\checkmark$&$\checkmark$&0	&$\checkmark$&0	&0\\
$J^{\mathrm{(AS)}59}_{ij}$&0	&$\checkmark$&0	&0	&0	&0	&0	&$\checkmark$&0	&0	&$\checkmark$&0	&0	&$\checkmark$&$\checkmark$&0	&0\\
$J^{\mathrm{(AS)}67}_{ij}$&0	&0	&0	&0	&$\checkmark$&0	&$\checkmark$&0	&0	&0	&0	&0	&$\checkmark$&$\checkmark$&$\checkmark$&0	&0\\
$J^{\mathrm{(AS)}68}_{ij}$&0	&0	&0	&0	&$\checkmark$&0	&$\checkmark$&0	&0	&0	&0	&0	&$\checkmark$&$\checkmark$&$\checkmark$&0	&0\\
$J^{\mathrm{(AS)}69}_{ij}$&0	&0	&0	&$\checkmark$&0	&0	&$\checkmark$&0	&0	&0	&$\checkmark$&$\checkmark$&0	&0	&$\checkmark$&0	&0\\
$J^{\mathrm{(AS)}78}_{ij}$&0	&0	&0	&$\checkmark$&0	&0	&$\checkmark$&0	&0	&0	&$\checkmark$&$\checkmark$&0	&0	&$\checkmark$&0	&0\\
$J^{\mathrm{(AS)}79}_{ij}$&0	&0	&0	&0	&$\checkmark$&0	&$\checkmark$&0	&0	&0	&0	&0	&$\checkmark$&$\checkmark$&$\checkmark$&0	&0\\
$J^{\mathrm{(AS)}89}_{ij}$&0	&0	&0	&0	&$\checkmark$&0	&$\checkmark$&0	&0	&0	&0	&0	&$\checkmark$&$\checkmark$&$\checkmark$&0	&0
\\ \hline
\# & 0 &  5 & 5 & 2 & 6 & 0 & 8 & 10 & 0 & 0 & 7 & 7 & 11 & 11 & 18 & 0 & 0
\\ \hline \hline 
\end{tabular}
}
\end{table*}

\begin{table*}
\centering
\caption{
\label{tab:hexagonal_H_S2}
Classification of the symmetric hexadecapole interactions with the hexagonal bases for the $[001]$ bond.
The row ``\#" represents the number of the independent interaction parameters except for the diagonal components.
}
\scalebox{0.85}{
\begin{tabular}{ccccccccccccccccccc}
\hline \hline
& $6/mmm$ & $\bar{6}m2$ & $\bar{6}2m$ & $6mm$ & $622$ & $6/m..$ & $6..$ & $\bar{6}..$ & $\bar{3}m1$ & $\bar{3}1m$ & $3m1$ & $31m$ & $321$ & $312$ & $3..$  & $\bar{3}..$ & $mmm$   \\ \hline
$J^{\mathrm{(S)}12}_{ij}$&0	&0	&0	&0	&0	&0	&0	&0	&$\checkmark$&0	&$\checkmark$&0	&$\checkmark$&0	&$\checkmark$&$\checkmark$&0\\
$J^{\mathrm{(S)}13}_{ij}$&0	&0	&0	&0	&0	&0	&0	&0	&0	&$\checkmark$&0	&$\checkmark$&0	&$\checkmark$&$\checkmark$&$\checkmark$&0\\
$J^{\mathrm{(S)}14}_{ij}$&0	&0	&0	&0	&0	&0	&0	&0	&0	&0	&0	&0	&0	&0	&0	&0	&0\\
$J^{\mathrm{(S)}15}_{ij}$&0	&0	&0	&0	&0	&0	&0	&0	&0	&0	&0	&0	&0	&0	&0	&0	&0\\
$J^{\mathrm{(S)}16}_{ij}$&0	&0	&0	&0	&0	&0	&0	&0	&0	&0	&0	&0	&0	&0	&0	&0	&$\checkmark$\\
$J^{\mathrm{(S)}17}_{ij}$&0	&0	&0	&0	&0	&0	&0	&0	&0	&0	&0	&0	&0	&0	&0	&0	&0\\
$J^{\mathrm{(S)}18}_{ij}$&0	&0	&0	&0	&0	&0	&0	&0	&0	&0	&0	&0	&0	&0	&0	&0	&0\\
$J^{\mathrm{(S)}19}_{ij}$&0	&0	&0	&0	&0	&0	&0	&0	&0	&0	&0	&0	&0	&0	&0	&0	&$\checkmark$\\
$J^{\mathrm{(S)}23}_{ij}$&0	&0	&0	&0	&0	&$\checkmark$&$\checkmark$&$\checkmark$&0	&0	&0	&0	&0	&0	&$\checkmark$&$\checkmark$&0\\
$J^{\mathrm{(S)}24}_{ij}$&0	&0	&0	&0	&0	&0	&0	&0	&0	&0	&0	&0	&0	&0	&0	&0	&0\\
$J^{\mathrm{(S)}25}_{ij}$&0	&0	&0	&0	&0	&0	&0	&0	&0	&0	&0	&0	&0	&0	&0	&0	&$\checkmark$\\
$J^{\mathrm{(S)}26}_{ij}$&0	&0	&0	&0	&0	&0	&0	&0	&0	&0	&0	&0	&0	&0	&0	&0	&0\\
$J^{\mathrm{(S)}27}_{ij}$&0	&0	&0	&0	&0	&0	&0	&0	&0	&0	&0	&0	&0	&0	&0	&0	&0\\
$J^{\mathrm{(S)}28}_{ij}$&0	&0	&0	&0	&0	&0	&0	&0	&0	&0	&0	&0	&0	&0	&0	&0	&0\\
$J^{\mathrm{(S)}29}_{ij}$&0	&0	&0	&0	&0	&0	&0	&0	&0	&0	&0	&0	&0	&0	&0	&0	&0\\
$J^{\mathrm{(S)}34}_{ij}$&0	&0	&0	&0	&0	&0	&0	&0	&0	&0	&0	&0	&0	&0	&0	&0	&$\checkmark$\\
$J^{\mathrm{(S)}35}_{ij}$&0	&0	&0	&0	&0	&0	&0	&0	&0	&0	&0	&0	&0	&0	&0	&0	&0\\
$J^{\mathrm{(S)}36}_{ij}$&0	&0	&0	&0	&0	&0	&0	&0	&0	&0	&0	&0	&0	&0	&0	&0	&0\\
$J^{\mathrm{(S)}37}_{ij}$&0	&0	&0	&0	&0	&0	&0	&0	&0	&0	&0	&0	&0	&0	&0	&0	&0\\
$J^{\mathrm{(S)}38}_{ij}$&0	&0	&0	&0	&0	&0	&0	&0	&0	&0	&0	&0	&0	&0	&0	&0	&0\\
$J^{\mathrm{(S)}39}_{ij}$&0	&0	&0	&0	&0	&0	&0	&0	&0	&0	&0	&0	&0	&0	&0	&0	&0\\
$J^{\mathrm{(S)}45}_{ij}$&0	&0	&0	&0	&0	&0	&0	&0	&0	&0	&0	&0	&0	&0	&0	&0	&0\\
$J^{\mathrm{(S)}46}_{ij}$&0	&0	&0	&0	&0	&0	&0	&0	&0	&$\checkmark$&0	&$\checkmark$&0	&$\checkmark$&$\checkmark$&$\checkmark$&0\\
$J^{\mathrm{(S)}47}_{ij}$&0	&0	&0	&0	&0	&0	&0	&0	&$\checkmark$&0	&$\checkmark$&0	&$\checkmark$&0	&$\checkmark$&$\checkmark$&0\\
$J^{\mathrm{(S)}48}_{ij}$&0	&0	&0	&0	&0	&0	&0	&0	&$\checkmark$&0	&$\checkmark$&0	&$\checkmark$&0	&$\checkmark$&$\checkmark$&0\\
$J^{\mathrm{(S)}49}_{ij}$&0	&0	&0	&0	&0	&0	&0	&0	&0	&$\checkmark$&0	&$\checkmark$&0	&$\checkmark$&$\checkmark$&$\checkmark$&0\\
$J^{\mathrm{(S)}56}_{ij}$&0	&0	&0	&0	&0	&0	&0	&0	&$\checkmark$&0	&$\checkmark$&0	&$\checkmark$&0	&$\checkmark$&$\checkmark$&0\\
$J^{\mathrm{(S)}57}_{ij}$&0	&0	&0	&0	&0	&0	&0	&0	&0	&$\checkmark$&0	&$\checkmark$&0	&$\checkmark$&$\checkmark$&$\checkmark$&0\\
$J^{\mathrm{(S)}58}_{ij}$&0	&0	&0	&0	&0	&0	&0	&0	&0	&$\checkmark$&0	&$\checkmark$&0	&$\checkmark$&$\checkmark$&$\checkmark$&0\\
$J^{\mathrm{(S)}59}_{ij}$&0	&0	&0	&0	&0	&0	&0	&0	&$\checkmark$&0	&$\checkmark$&0	&$\checkmark$&0	&$\checkmark$&$\checkmark$&0\\
$J^{\mathrm{(S)}67}_{ij}$&0	&0	&0	&0	&0	&0	&0	&0	&0	&0	&0	&0	&0	&0	&0	&0	&0\\
$J^{\mathrm{(S)}68}_{ij}$&0	&0	&0	&0	&0	&$\checkmark$&$\checkmark$&$\checkmark$&0	&0	&0	&0	&0	&0	&$\checkmark$&$\checkmark$&0\\
$J^{\mathrm{(S)}69}_{ij}$&$\checkmark$&$\checkmark$&$\checkmark$&$\checkmark$&$\checkmark$&$\checkmark$&$\checkmark$&$\checkmark$&$\checkmark$&$\checkmark$&$\checkmark$&$\checkmark$&$\checkmark$&$\checkmark$&$\checkmark$&$\checkmark$&$\checkmark$\\
$J^{\mathrm{(S)}78}_{ij}$&$\checkmark$&$\checkmark$&$\checkmark$&$\checkmark$&$\checkmark$&$\checkmark$&$\checkmark$&$\checkmark$&$\checkmark$&$\checkmark$&$\checkmark$&$\checkmark$&$\checkmark$&$\checkmark$&$\checkmark$&$\checkmark$&$\checkmark$\\
$J^{\mathrm{(S)}79}_{ij}$&0	&0	&0	&0	&0	&$\checkmark$&$\checkmark$&$\checkmark$&0	&0	&0	&0	&0	&0	&$\checkmark$&$\checkmark$&0\\
$J^{\mathrm{(S)}89}_{ij}$&0	&0	&0	&0	&0	&0	&0	&0	&0	&0	&0	&0	&0	&0	&0	&0	&0
\\ \hline
\# & 2 &  2 & 2 & 2 & 2 & 5 & 5 & 5 & 7 & 7 & 7 & 7 & 7 & 7 & 15 & 15 & 6
\\ \hline \hline 
\end{tabular}
}
\end{table*}

We show the classification of the antisymmetric and symmetric hexadecapole interactions with the hexagonal basis for the [001] bond in Tables~\ref{tab:hexagonal_H_AS2} and \ref{tab:hexagonal_H_S2}, respectively.

\section{Multipole Interactions under Cubic Bases for the [001] Bond}
\label{sec: app3}

By using the cubic bases in Table~\ref{tab:cubic_basis}, we classify the multipole interactions for the [001] bond with the point group symmetries $4/mmm$, $\bar{4}m2$, $\bar{4}2m$, $4mm$, $422$, $4/m..$,  $4..$, $\bar{4}..$, and $mmm$.

\subsection{Quadrupole Interaction}

\begin{table}
\centering
\caption{
\label{tab:cubic_Q_AS2}
Classification of the antisymmetric quadrupole interactions with the cubic bases for the $[001]$ bond.
The row ``\#" represents the number of the independent interaction parameters.
}
\scalebox{0.8}{
\begin{tabular}{cccccccccc}
\hline \hline 
& $4/mmm$ & $\bar{4}m2$ & $\bar{4}2m$ & $4mm$ & $422$ & $4/m..$ & $4..$ & $\bar{4}..$ & $mmm$   \\ \hline
$J^{\mathrm{(AS)}12}_{ij}$&0	&0	&0	&0	&0	&0	&0	&0	&0\\
$J^{\mathrm{(AS)}13}_{ij}$&0	&0	&0	&0	&0	&0	&0	&0	&0\\
$J^{\mathrm{(AS)}14}_{ij}$&0	&0	&0	&0	&0	&0	&0	&0	&0\\
$J^{\mathrm{(AS)}15}_{ij}$&0	&0	&0	&0	&0	&0	&0	&0	&0\\
$J^{\mathrm{(AS)}23}_{ij}$&0	&0	&$\checkmark$&0	&$\checkmark$&0	&$\checkmark$&$\checkmark$&0\\
$J^{\mathrm{(AS)}24}_{ij}$&0	&0	&0	&0	&0	&0	&0	&0	&0\\
$J^{\mathrm{(AS)}25}_{ij}$&0	&0	&0	&0	&0	&0	&0	&0	&0\\
$J^{\mathrm{(AS)}34}_{ij}$&0	&0	&0	&0	&0	&0	&0	&0	&0\\
$J^{\mathrm{(AS)}35}_{ij}$&0	&0	&0	&0	&0	&0	&0	&0	&0\\
$J^{\mathrm{(AS)}45}_{ij}$&0	&0	&$\checkmark$&0	&$\checkmark$&0	&$\checkmark$&$\checkmark$&0
\\ \hline
\# & 0 &  0 & 2 & 0 & 2 & 0 & 2 & 2 & 0  
\\ \hline \hline 
\end{tabular}
}
\end{table}

\begin{table}
\centering
\caption{
\label{tab:cubic_Q_S2}
Classification of the symmetric quadrupole interactions with the cubic bases for the $[001]$ bond.
The row ``\#" represents the number of the independent interaction parameters except for the diagonal components.
}
\scalebox{0.8}{
\begin{tabular}{cccccccccc}
\hline \hline 
& $4/mmm$ & $\bar{4}m2$ & $\bar{4}2m$ & $4mm$ & $422$ & $4/m..$ & $4..$ & $\bar{4}..$ & $mmm$   \\ \hline
$J^{\mathrm{(S)}12}_{ij}$&0	&0	&0	&0	&0	&0	&0	&0	&$\checkmark$\\
$J^{\mathrm{(S)}13}_{ij}$&0	&0	&0	&0	&0	&0	&0	&0	&0\\
$J^{\mathrm{(S)}14}_{ij}$&0	&0	&0	&0	&0	&0	&0	&0	&0\\
$J^{\mathrm{(S)}15}_{ij}$&0	&0	&0	&0	&0	&0	&0	&0	&0\\
$J^{\mathrm{(S)}23}_{ij}$&0	&0	&0	&0	&0	&$\checkmark$ &$\checkmark$ &$\checkmark$ &0\\
$J^{\mathrm{(S)}24}_{ij}$&0	&0	&0	&0	&0	&0	&0	&0	&0\\
$J^{\mathrm{(S)}25}_{ij}$&0	&0	&0	&0	&0	&0	&0	&0	&0\\
$J^{\mathrm{(S)}34}_{ij}$&0	&0	&0	&0	&0	&0	&0	&0	&0\\
$J^{\mathrm{(S)}35}_{ij}$&0	&0	&0	&0	&0	&0	&0	&0	&0\\
$J^{\mathrm{(S)}45}_{ij}$&0	&0	&0	&0	&0	&0	&0	&0	&0
\\ \hline
\# & 0 &  0 & 0 & 0 & 0 & 1 & 1 & 1 & 1  
\\ \hline \hline 
\end{tabular}
}
\end{table}

We show the classification of the antisymmetric and symmetric quadrupole interactions with the cubic basis for the [001] bond in Tables~\ref{tab:cubic_Q_AS2} and \ref{tab:cubic_Q_S2}, respectively.

\subsection{Octupole Interaction}

\begin{table}
\centering
\caption{
\label{tab:cubic_O_AS2}
Classification of the antisymmetric octupole interactions with the cubic bases for the $[001]$ bond.
The row ``\#" represents the number of the independent interaction parameters.
}
\scalebox{0.8}{
\begin{tabular}{cccccccccc}
\hline \hline 
& $4/mmm$ & $\bar{4}m2$ & $\bar{4}2m$ & $4mm$ & $422$ & $4/m..$ & $4..$ & $\bar{4}..$ & $mmm$   \\ \hline
$J^{\mathrm{(AS)}12}_{ij}$&0	&0	&0	&0	&0	&0	&0	&0	&0\\
$J^{\mathrm{(AS)}13}_{ij}$&0	&0	&0	&0	&0	&0	&0	&0	&0\\
$J^{\mathrm{(AS)}14}_{ij}$&0	&0	&0	&0	&0	&0	&0	&0	&0\\
$J^{\mathrm{(AS)}15}_{ij}$&0	&0	&0	&0	&0	&0	&0	&0	&0\\
$J^{\mathrm{(AS)}16}_{ij}$&0	&0	&0	&0	&0	&0	&0	&0	&0\\
$J^{\mathrm{(AS)}17}_{ij}$&0	&0	&0	&0	&0	&0	&0	&0	&0\\
$J^{\mathrm{(AS)}23}_{ij}$&0	&0	&$\checkmark$&0	&$\checkmark$&0	&$\checkmark$&$\checkmark$&0\\
$J^{\mathrm{(AS)}24}_{ij}$&0	&0	&0	&0	&0	&0	&0	&0	&0\\
$J^{\mathrm{(AS)}25}_{ij}$&0	&0	&0	&0	&0	&0	&0	&0	&0\\
$J^{\mathrm{(AS)}26}_{ij}$&0	&0	&0	&0	&0	&0	&0	&0	&0\\
$J^{\mathrm{(AS)}27}_{ij}$&0	&0	&0	&0	&0	&0	&0	&0	&0\\
$J^{\mathrm{(AS)}34}_{ij}$&0	&0	&0	&0	&0	&0	&0	&0	&0\\
$J^{\mathrm{(AS)}35}_{ij}$&0	&0	&0	&0	&0	&0	&0	&0	&0\\
$J^{\mathrm{(AS)}36}_{ij}$&0	&0	&0	&0	&0	&0	&0	&0	&0\\
$J^{\mathrm{(AS)}37}_{ij}$&0	&0	&0	&0	&0	&0	&0	&0	&0\\
$J^{\mathrm{(AS)}45}_{ij}$&0	&0	&$\checkmark$&0	&$\checkmark$&0	&$\checkmark$&$\checkmark$&0\\
$J^{\mathrm{(AS)}46}_{ij}$&0	&0	&$\checkmark$&0	&$\checkmark$&0	&$\checkmark$&$\checkmark$&0\\
$J^{\mathrm{(AS)}47}_{ij}$&0	&$\checkmark$&0	&$\checkmark$&0	&0	&$\checkmark$&$\checkmark$&0\\
$J^{\mathrm{(AS)}56}_{ij}$&0	&$\checkmark$&0	&$\checkmark$&0	&0	&$\checkmark$&$\checkmark$&0\\
$J^{\mathrm{(AS)}57}_{ij}$&0	&0	&$\checkmark$&0	&$\checkmark$&0	&$\checkmark$&$\checkmark$&0\\
$J^{\mathrm{(AS)}67}_{ij}$&0	&0	&$\checkmark$&0	&$\checkmark$&0	&$\checkmark$&$\checkmark$&0
\\ \hline
\# & 0 &  2 & 5 & 2 & 5 & 0 & 7 & 7 & 0  
\\ \hline \hline 
\end{tabular}
}
\end{table}

\begin{table}
\centering
\caption{
\label{tab:cubic_O_S2}
Classification of the symmetric octupole interactions with the cubic bases for the $[001]$ bond.
The row ``\#" represents the number of the independent interaction parameters except for the diagonal components.
}
\scalebox{0.8}{
\begin{tabular}{cccccccccc}
\hline \hline 
& $4/mmm$ & $\bar{4}m2$ & $\bar{4}2m$ & $4mm$ & $422$ & $4/m..$ & $4..$ & $\bar{4}..$ & $mmm$   \\ \hline
$J^{\mathrm{(S)}12}_{ij}$&0	&0	&0	&0	&0	&0	&0	&0	&0\\
$J^{\mathrm{(S)}13}_{ij}$&0	&0	&0	&0	&0	&0	&0	&0	&$\checkmark$\\
$J^{\mathrm{(S)}14}_{ij}$&0	&0	&0	&0	&0	&0	&0	&0	&0\\
$J^{\mathrm{(S)}15}_{ij}$&0	&0	&0	&0	&0	&0	&0	&0	&0\\
$J^{\mathrm{(S)}16}_{ij}$&0	&0	&0	&0	&0	&0	&0	&0	&0\\
$J^{\mathrm{(S)}17}_{ij}$&0	&0	&0	&0	&0	&0	&0	&0	&0\\
$J^{\mathrm{(S)}23}_{ij}$&0	&0	&0	&0	&0	&$\checkmark$&$\checkmark$&$\checkmark$&0\\
$J^{\mathrm{(S)}24}_{ij}$&0	&0	&0	&0	&0	&0	&0	&0	&0\\
$J^{\mathrm{(S)}25}_{ij}$&0	&0	&0	&0	&0	&0	&0	&0	&0\\
$J^{\mathrm{(S)}26}_{ij}$&0	&0	&0	&0	&0	&0	&0	&0	&0\\
$J^{\mathrm{(S)}27}_{ij}$&0	&0	&0	&0	&0	&0	&0	&0	&0\\
$J^{\mathrm{(S)}34}_{ij}$&0	&0	&0	&0	&0	&0	&0	&0	&0\\
$J^{\mathrm{(S)}35}_{ij}$&0	&0	&0	&0	&0	&0	&0	&0	&0\\
$J^{\mathrm{(S)}36}_{ij}$&0	&0	&0	&0	&0	&0	&0	&0	&0\\
$J^{\mathrm{(S)}37}_{ij}$&0	&0	&0	&0	&0	&0	&0	&0	&0\\
$J^{\mathrm{(S)}45}_{ij}$&0	&0	&0	&0	&0	&0	&0	&0	&0\\
$J^{\mathrm{(S)}46}_{ij}$&0	&0	&0	&0	&0	&$\checkmark$&$\checkmark$&$\checkmark$&0\\
$J^{\mathrm{(S)}47}_{ij}$&$\checkmark$&$\checkmark$&$\checkmark$&$\checkmark$&$\checkmark$&$\checkmark$&$\checkmark$&$\checkmark$ &$\checkmark$\\
$J^{\mathrm{(S)}56}_{ij}$&$\checkmark$&$\checkmark$&$\checkmark$&$\checkmark$&$\checkmark$&$\checkmark$&$\checkmark$&$\checkmark$ &$\checkmark$\\
$J^{\mathrm{(S)}57}_{ij}$&0	&0	&0	&0	&0	&$\checkmark$&$\checkmark$&$\checkmark$&0\\
$J^{\mathrm{(S)}67}_{ij}$&0	&0	&0	&0	&0	&0	&0	&0	&0
\\ \hline
\# & 2 &  2 & 2 & 2 & 2 & 5 & 5 & 5 & 3  
\\ \hline \hline 
\end{tabular}
}
\end{table}

We show the classification of the antisymmetric and symmetric octupole interactions with the cubic basis for the [001] bond in Tables~\ref{tab:cubic_O_AS2} and \ref{tab:cubic_O_S2}, respectively.

\subsection{Hexadecapole Interaction}

\begin{table}
\centering
\caption{
\label{tab:cubic_H_AS2}
Classification of the antisymmetric hexadecapole interactions with the cubic bases for the $[001]$ bond.
The row ``\#" represents the number of the independent interaction parameters.
}
\scalebox{0.8}{
\begin{tabular}{cccccccccc}
\hline \hline 
& $4/mmm$ & $\bar{4}m2$ & $\bar{4}2m$ & $4mm$ & $422$ & $4/m..$ & $4..$ & $\bar{4}..$ & $mmm$   \\ \hline
$J^{\mathrm{(AS)}12}_{ij}$&0	&$\checkmark$&0	&$\checkmark$&0	&0	&$\checkmark$&$\checkmark$&0\\
$J^{\mathrm{(AS)}13}_{ij}$&0	&0	&$\checkmark$&0	&$\checkmark$&0	&$\checkmark$&$\checkmark$&0\\
$J^{\mathrm{(AS)}14}_{ij}$&0	&0	&0	&0	&0	&0	&0	&0	&0\\
$J^{\mathrm{(AS)}15}_{ij}$&0	&0	&0	&0	&0	&0	&0	&0	&0\\
$J^{\mathrm{(AS)}16}_{ij}$&0	&0	&0	&0	&0	&0	&0	&0	&0\\
$J^{\mathrm{(AS)}17}_{ij}$&0	&0	&0	&0	&0	&0	&0	&0	&0\\
$J^{\mathrm{(AS)}18}_{ij}$&0	&0	&0	&0	&0	&0	&0	&0	&0\\
$J^{\mathrm{(AS)}19}_{ij}$&0	&0	&0	&0	&0	&0	&0	&0	&0\\
$J^{\mathrm{(AS)}23}_{ij}$&0	&0	&$\checkmark$&0	&$\checkmark$&0	&$\checkmark$&$\checkmark$&0\\
$J^{\mathrm{(AS)}24}_{ij}$&0	&0	&0	&0	&0	&0	&0	&0	&0\\
$J^{\mathrm{(AS)}25}_{ij}$&0	&0	&0	&0	&0	&0	&0	&0	&0\\
$J^{\mathrm{(AS)}26}_{ij}$&0	&0	&0	&0	&0	&0	&0	&0	&0\\
$J^{\mathrm{(AS)}27}_{ij}$&0	&0	&0	&0	&0	&0	&0	&0	&0\\
$J^{\mathrm{(AS)}28}_{ij}$&0	&0	&0	&0	&0	&0	&0	&0	&0\\
$J^{\mathrm{(AS)}29}_{ij}$&0	&0	&0	&0	&0	&0	&0	&0	&0\\
$J^{\mathrm{(AS)}34}_{ij}$&0	&0	&0	&0	&0	&0	&0	&0	&0\\
$J^{\mathrm{(AS)}35}_{ij}$&0	&0	&0	&0	&0	&0	&0	&0	&0\\
$J^{\mathrm{(AS)}36}_{ij}$&0	&0	&0	&0	&0	&0	&0	&0	&0\\
$J^{\mathrm{(AS)}37}_{ij}$&0	&0	&0	&0	&0	&0	&0	&0	&0\\
$J^{\mathrm{(AS)}38}_{ij}$&0	&0	&0	&0	&0	&0	&0	&0	&0\\
$J^{\mathrm{(AS)}39}_{ij}$&0	&0	&0	&0	&0	&0	&0	&0	&0\\
$J^{\mathrm{(AS)}45}_{ij}$&0	&0	&$\checkmark$&0	&$\checkmark$&0	&$\checkmark$&$\checkmark$&0\\
$J^{\mathrm{(AS)}46}_{ij}$&0	&0	&0	&0	&0	&0	&0	&0	&0\\
$J^{\mathrm{(AS)}47}_{ij}$&0	&0	&0	&0	&0	&0	&0	&0	&0\\
$J^{\mathrm{(AS)}48}_{ij}$&0	&0	&0	&0	&0	&0	&0	&0	&0\\
$J^{\mathrm{(AS)}49}_{ij}$&0	&0	&0	&0	&0	&0	&0	&0	&0\\
$J^{\mathrm{(AS)}56}_{ij}$&0	&0	&0	&0	&0	&0	&0	&0	&0\\
$J^{\mathrm{(AS)}57}_{ij}$&0	&0	&0	&0	&0	&0	&0	&0	&0\\
$J^{\mathrm{(AS)}58}_{ij}$&0	&0	&0	&0	&0	&0	&0	&0	&0\\
$J^{\mathrm{(AS)}59}_{ij}$&0	&0	&0	&0	&0	&0	&0	&0	&0\\
$J^{\mathrm{(AS)}67}_{ij}$&0	&0	&$\checkmark$&0	&$\checkmark$&0	&$\checkmark$&$\checkmark$&0\\
$J^{\mathrm{(AS)}68}_{ij}$&0	&0	&$\checkmark$&0	&$\checkmark$&0	&$\checkmark$&$\checkmark$&0\\
$J^{\mathrm{(AS)}69}_{ij}$&0	&$\checkmark$&0	&$\checkmark$&0	&0	&$\checkmark$&$\checkmark$&0\\
$J^{\mathrm{(AS)}78}_{ij}$&0	&$\checkmark$&0	&$\checkmark$&0	&0	&$\checkmark$&$\checkmark$&0\\
$J^{\mathrm{(AS)}79}_{ij}$&0	&0	&$\checkmark$&0	&$\checkmark$&0	&$\checkmark$&$\checkmark$&0\\
$J^{\mathrm{(AS)}89}_{ij}$&0	&0	&$\checkmark$&0	&$\checkmark$&0	&$\checkmark$&$\checkmark$&0
\\ \hline
\# & 0 &  3 & 7 & 3 & 7 & 0 & 10 & 10 & 0  
\\ \hline \hline 
\end{tabular}
}
\end{table}

\begin{table}
\centering
\caption{
\label{tab:cubic_H_S2}
Classification of the symmetric hexadecapole interactions with the cubic bases for the $[001]$ bond.
The row ``\#" represents the number of the independent interaction parameters except for the diagonal components.
}
\scalebox{0.8}{
\begin{tabular}{cccccccccc}
\hline \hline 
& $4/mmm$ & $\bar{4}m2$ & $\bar{4}2m$ & $4mm$ & $422$ & $4/m..$ & $4..$ & $\bar{4}..$ & $mmm$   \\ \hline
$J^{\mathrm{(S)}12}_{ij}$&$\checkmark$&$\checkmark$&$\checkmark$&$\checkmark$&$\checkmark$&$\checkmark$&$\checkmark$&$\checkmark$&$\checkmark$\\
$J^{\mathrm{(S)}13}_{ij}$&0	&0	&0	&0	&0	&$\checkmark$&$\checkmark$&$\checkmark$&0\\
$J^{\mathrm{(S)}14}_{ij}$&0	&0	&0	&0	&0	&0	&0	&0	&$\checkmark$\\
$J^{\mathrm{(S)}15}_{ij}$&0	&0	&0	&0	&0	&0	&0	&0	&0\\
$J^{\mathrm{(S)}16}_{ij}$&0	&0	&0	&0	&0	&0	&0	&0	&0\\
$J^{\mathrm{(S)}17}_{ij}$&0	&0	&0	&0	&0	&0	&0	&0	&0\\
$J^{\mathrm{(S)}18}_{ij}$&0	&0	&0	&0	&0	&0	&0	&0	&0\\
$J^{\mathrm{(S)}19}_{ij}$&0	&0	&0	&0	&0	&0	&0	&0	&0\\
$J^{\mathrm{(S)}23}_{ij}$&0	&0	&0	&0	&0	&$\checkmark$&$\checkmark$&$\checkmark$&0\\
$J^{\mathrm{(S)}24}_{ij}$&0	&0	&0	&0	&0	&0	&0	&0	&$\checkmark$\\
$J^{\mathrm{(S)}25}_{ij}$&0	&0	&0	&0	&0	&0	&0	&0	&0\\
$J^{\mathrm{(S)}26}_{ij}$&0	&0	&0	&0	&0	&0	&0	&0	&0\\
$J^{\mathrm{(S)}27}_{ij}$&0	&0	&0	&0	&0	&0	&0	&0	&0\\
$J^{\mathrm{(S)}28}_{ij}$&0	&0	&0	&0	&0	&0	&0	&0	&0\\
$J^{\mathrm{(S)}29}_{ij}$&0	&0	&0	&0	&0	&0	&0	&0	&0\\
$J^{\mathrm{(S)}34}_{ij}$&0	&0	&0	&0	&0	&0	&0	&0	&0\\
$J^{\mathrm{(S)}35}_{ij}$&0	&0	&0	&0	&0	&0	&0	&0	&$\checkmark$\\
$J^{\mathrm{(S)}36}_{ij}$&0	&0	&0	&0	&0	&0	&0	&0	&0\\
$J^{\mathrm{(S)}37}_{ij}$&0	&0	&0	&0	&0	&0	&0	&0	&0\\
$J^{\mathrm{(S)}38}_{ij}$&0	&0	&0	&0	&0	&0	&0	&0	&0\\
$J^{\mathrm{(S)}39}_{ij}$&0	&0	&0	&0	&0	&0	&0	&0	&0\\
$J^{\mathrm{(S)}45}_{ij}$&0	&0	&0	&0	&0	&$\checkmark$&$\checkmark$&$\checkmark$&0\\
$J^{\mathrm{(S)}46}_{ij}$&0	&0	&0	&0	&0	&0	&0	&0	&0\\
$J^{\mathrm{(S)}47}_{ij}$&0	&0	&0	&0	&0	&0	&0	&0	&0\\
$J^{\mathrm{(S)}48}_{ij}$&0	&0	&0	&0	&0	&0	&0	&0	&0\\
$J^{\mathrm{(S)}49}_{ij}$&0	&0	&0	&0	&0	&0	&0	&0	&0\\
$J^{\mathrm{(S)}56}_{ij}$&0	&0	&0	&0	&0	&0	&0	&0	&0\\
$J^{\mathrm{(S)}57}_{ij}$&0	&0	&0	&0	&0	&0	&0	&0	&0\\
$J^{\mathrm{(S)}58}_{ij}$&0	&0	&0	&0	&0	&0	&0	&0	&0\\
$J^{\mathrm{(S)}59}_{ij}$&0	&0	&0	&0	&0	&0	&0	&0	&0\\
$J^{\mathrm{(S)}67}_{ij}$&0	&0	&0	&0	&0	&0	&0	&0	&0\\
$J^{\mathrm{(S)}68}_{ij}$&0	&0	&0	&0	&0	&$\checkmark$&$\checkmark$&$\checkmark$&0\\
$J^{\mathrm{(S)}69}_{ij}$&$\checkmark$&$\checkmark$&$\checkmark$&$\checkmark$&$\checkmark$&$\checkmark$&$\checkmark$&$\checkmark$&$\checkmark$\\
$J^{\mathrm{(S)}78}_{ij}$&$\checkmark$&$\checkmark$&$\checkmark$&$\checkmark$&$\checkmark$&$\checkmark$&$\checkmark$&$\checkmark$&$\checkmark$\\
$J^{\mathrm{(S)}79}_{ij}$&0	&0	&0	&0	&0	&$\checkmark$&$\checkmark$&$\checkmark$&0\\
$J^{\mathrm{(S)}89}_{ij}$&0	&0	&0	&0	&0	&0	&0	&0	&0
\\ \hline
\# & 3 &  3 & 3 & 3 & 3 & 8 & 8 & 8 & 6  
\\ \hline \hline 
\end{tabular}
}
\end{table}

We show the classification of the antisymmetric and symmetric hexadecapole interactions with the cubic basis for the [001] bond in Tables~\ref{tab:cubic_H_AS2} and \ref{tab:cubic_H_S2}, respectively. 

\bibliographystyle{jpsj}
\bibliography{ref.bib}

\end{document}